\documentclass[prd,showpacs,preprintnumbers,amsmath,amssymb,eqsecnum]{revtex4}
\usepackage{amsmath,amssymb,graphicx}
\usepackage{epsf}
\usepackage{epstopdf}
\usepackage {amssymb}
\usepackage{wick}

\newcommand{\nc}{\newcommand}
\nc{\ba}{\begin{eqnarray}}
\nc{\ea}{\end{eqnarray}}
\newcommand\be{\begin{equation}}
\newcommand\ee{\end{equation}}

\newcommand{\bmk}{\mathbf{k}}
\newcommand{\bmq}{\mathbf{q}}
\newcommand{\bmp}{\mathbf{p}}
\newcommand{\delpsi}{\left(\delta\chi^2\right)}
\newcommand{\bmx}{\mathbf{x}}

\nc{\x}{{\bf{x}}}
\newcommand{\calR}{{\cal R}}
\newcommand{\calN}{{\cal N}}

\begin{document}

\title{ Local Features with Large Spiky non-Gaussianities during Inflation }

\author{Ali Akbar Abolhasani$^{1}$}
\email{abolhasani-AT-mail.ipm.ir}
\author{Hassan Firouzjahi$^{1}$}
\email{firouz-AT-mail.ipm.ir}
\author{Shahram Khosravi$^{2}$}
\email{khosravi-AT-mail.ipm.ir}
\author{Misao Sasaki$^{3}$}
\email{misao-AT-yukawa.kyoto-u.ac.jp}
\affiliation{$^1$School of Physics, Institute for Research in 
Fundamental Sciences (IPM),
P. O. Box 19395-5531,
Tehran, Iran}
\affiliation{$^2$Physics Department, Faculty of Science, 
Tarbiat Moallem University, Tehran, Iran, and  School of Astronomy, 
Institute for Research in Fundamental Sciences(IPM), Tehran, Iran. }
\affiliation{$^3$Yukawa Institute for theoretical Physics,
 Kyoto University, Kyoto 606-8502, Japan}

\date{\today}

\begin{abstract}
\vspace{0.3cm}
We provide a dynamical mechanism to generate  localized features
during  inflation.
The local feature is due to a sharp waterfall phase transition 
which is coupled to the inflaton field.  The key effect is 
the contributions of waterfall quantum fluctuations which induce
 a sharp peak on the curvature perturbation which can be as large
 as the background curvature perturbation from inflaton field.  
Due to non-Gaussian nature of waterfall quantum fluctuations
 a large spike non-Gaussianity is produced which is narrowly 
peaked at modes which leave the Hubble radius at the time of
phase transition. The large localized peaks in power spectrum 
and bispectrum can have interesting consequences on CMB anisotropies.  

\vspace{0.3cm}

\end{abstract}

\preprint{YITP-12-16,   IPM/P-2012/011 }

\maketitle

\section{Introduction}
\label{sec:introduction}

Simplest inflationary scenarios predict almost scale invariant,
almost Gaussian and almost adiabatic perturbations and
the observed cosmic microwave background (CMB) anisotropies
are in very good agreements with these predictions~\cite{Komatsu:2010fb}.
However, with higher precision data expected to come from near 
future cosmological observations such as PLANCK, one hopes that
some deviations from the predictions of single-field inflation
may be observed. In particular a detection of non-Gaussianity
would exclude a large class of inflationary scenarios.

There have been many attempts in the literature to generalize 
local features during inflation which may have non-trivial predictions
 on the power spectrum and 
bispectrum~\cite{Starobinsky:1992ts, Leach:2001zf, Adams:2001vc, Gong:2005jr, Joy:2007na, Chen:2006xjb, Chen:2008wn, Chen:2011zf, Chen:2011tu, Hotchkiss:2009pj, Arroja:2011yu, Adshead:2011jq}. 
This is partly motivated from  glitches in the CMB angular power spectrum 
at $\ell \sim 20-40$. These local features may originate from models 
of high energy physics, particle creation or field annihilation during
 inflation~\cite{Romano:2008rr, Battefeld:2010rf, Firouzjahi:2010ga, Battefeld:2010vr, Barnaby:2009dd, Barnaby:2010ke, Biswas:2010si, Silverstein:2008sg, Flauger:2009ab, Flauger:2010ja, Bean:2008na}.

In many of these models local features are generated by temporal 
violations of slow-roll conditions during inflation. 
However, many of them are based on rather ad hoc mechanisms 
to violate slow-roll conditions. In this work, we provide a 
dynamical mechanism to generate local features in the power spectrum 
and bispectrum. The model consists of a simple chaotic inflation potential,
$m^2 \phi^2/2$, coupled to a heavy field $\chi$. After $\phi$ reaches
 a critical value $\phi=\phi_c(\gg M_P)$, $\chi$ becomes tachyonic and
a rapid waterfall phase transition takes place. In this sense the model
is similar to hybrid inflation \cite{Linde:1993cn, Copeland:1994vg}. 
However, the key difference is that 
the potential is not vacuum dominated but the chaotic type inflaton 
potential is the main driving source of inflation.
Furthermore, inflation continues even after the waterfall phase transition.

In this model, inflation may be divided into three stages.
At the first stage $\phi> \phi_c$, inflation proceeds 
as in chaotic inflation. The second stage at which $\phi\lesssim\phi_c$
is fairly short, of the order of the Hubble time, and 
$\chi$ becomes tachyonic and a waterfall phase transition 
takes place where $\chi$ settles to its global minimum. 
This causes a small change in the inflaton effective mass,
$m^2\to m_+^2=m^2 (1+ C)$ where $C<<1$.
Then the third stage proceeds as in chaotic inflation
again. In order to bring the local feature into the observable
windon of CMB, we demand that the phase transition takes place 
$50 - 60$ $e$-folds before the end of inflation.

The key point in our analysis is the evaluation of the waterfall 
quantum fluctuations and its contribution to the curvature 
perturbation on uniform density slices $\zeta$ \cite{Wands:2000dp}.
For studies of the waterfall contribution to $\zeta$, 
in the context of hybrid inflation, 
see \cite{Lyth:2010ch,Abolhasani:2010kr,Fonseca:2010nk,
Abolhasani:2011yp,Gong:2010zf,Lyth:2012yp,Lyth:2011kj,
Levasseur:2010rk,Barnaby:2006km,Mazumdar:2010sa,
Enqvist:2004ey,Enqvist:2004bk,Randall:1995dj,GarciaBellido:1996qt,
Clesse:2010iz,Martin:2011ib,Mulryne:2011ni,Avgoustidis:2011em,
Kodama:2011vs,Abolhasani:2010kn, Bugaev:2011qt, Bugaev:2011wy}.

 Due to the sharpness of the waterfall phase transition, the waterfall 
contribution in $\zeta$ is narrowly localized around the modes which 
leave the horizon at the time of waterfall transition.
 Furthermore, due to the intrinsically non-Gaussian nature of the
waterfall contribution, which is in the form of 
$\zeta \sim \delta \chi^2$, a large spiky bispectrum is generated.
We emphasize that this is a local dynamical effect intrinsic to 
the waterfall quantum fluctuations which are absent in other models,
hence is an observationally testable feature genuinely
specific to our model.

The rest of paper is organized as follows. In Section \ref{sec:model} 
we present our model and basic setup.
In Section \ref{sec:deltaN} we calculate $\zeta $ using the
$\delta N$ formalism.
In Section \ref{sec:bispectrum} we calculate the dynamical and intrinsic 
non-Gaussianities in our scenario followed by conclusion and 
discussions in Section \ref{sec:conclusion}. Technical 
details about the power spectrum of the waterfall quantum fluctuations 
and higher order $\delta N$ perturbations are deferred to appendices.

\section{The Model}
\label{sec:model}

In this section we present our setup in modeling local features 
during inflation.
In our picture, we consider the simplest inflationary model, 
the chaotic potential $m^2 \phi^2/2$, where
we assume the mass of inflaton undergoes a sudden dynamical 
change at $\phi=\phi_c$. We would like to see how a sudden small
 change in the inflaton mass can be modeled in a consistent
 dynamical way and then look for
its observational consequences in CMB. One of the best known 
dynamical mechanism for
inducing local feature during inflation is the idea of waterfall 
phase transition. For this purpose, we consider the following setup:
\ba
\label{potential}
V=  \frac{m^2}{2} \phi^2
 + \frac{\lambda}{4} \left( \chi^2 - \frac{M^2}{\lambda} \right)^2  
+ \frac{g^2}{2} \phi^2 \chi^2  \, ,
\ea
where $\phi$ is the inflaton field and $\chi$ is the waterfall 
field. Formally potential (\ref{potential}) is identical to
 hybrid inflation \cite{Linde:1993cn, Copeland:1994vg}.
However, as we mentioned above, in our picture inflation is
 mainly driven by the field $\phi$
so in our picture effectively inflation proceeds as in 
chaotic model with potential $V \simeq m_{eff}^2 \phi^2/2$ 
with some effective mass $m_{eff}$ which undergoes a
small but abrupt change at $\phi=\phi_c$. The waterfall field
 $\chi$ is employed to induce this change in mass.

In this model, like in chaotic inflation, inflation starts 
at $\phi=\phi_i \gg M_P$, with $M_P^2= (8 \pi G)^{-1}$ for $G$ 
being the Newton constant, so one can obtain 60 $e$-folds or
 more to solve the horizon and flatness problems.
 The waterfall field is very heavy during the first stage of 
inflation so it remains at its instantaneous minimum $\chi=0$. 
Once the inflaton field reaches the critical value 
$\phi=\phi_c \equiv M/g$, the waterfall field $\chi$ becomes
 tachyonic and quickly rolls down to its global minimum 
$\chi_{min}^2= M^2/\lambda$. 
The final stage of inflation after $\phi>\phi_c$ proceeds 
as in chaotic inflation again
but with the effective mass of the inflaton, $m_+$, given by
\ba
m_+^2 = m^2 + g^2 \langle \chi^2 \rangle = m^2 \left( 1+ C \right)
\ea 
where
\ba
C\equiv \frac{g^2 M^2}{\lambda m^2} \, .
\ea
In our model, we assume the change in inflaton mass is small,
$C \ll 1$. In order to bring this local feature into the 
observable scale, we assume that the short waterfall stage which lasts
for about an $e$-fold or so begins at around 55 $e$-folds before
the end of inflation.

It is instructive to look into different contributions to
the potential before the phase transition where
$V = m^2\phi^2/2 + M^4/4\lambda$.  
We have $V(\phi_c) = m^2\phi_c^2/2 (1+ C/2)$. 
Having $C \ll 1$ in our model corresponds to the assumption 
that the inflationary potential is dominated by the 
$m^2 \phi^2/2$ term. This is in contrast to the standard hybrid 
inflation where the potential is vacuum dominated 
corresponding to $C\gg 1$. 

Before we proceed further we pause to compare our scenario to 
the one studied in \cite{ Joy:2007na} where the potential is 
still given by Eq.~(\ref{potential}). However, in \cite{ Joy:2007na} 
they are interested in the limit where $C \gtrsim 1$ so 
inflation is mildly vacuum dominated. Hence their model is effectively
a single field model where the waterfall quantum fluctuations can
be neglected. In contrast, in the present paper we pay careful attention
to the dynamics of the waterfall phase transition and calculate 
rigorously the curvature perturbations induced from the 
waterfall quantum fluctuations. For this to happen, 
we require $\phi_c \gtrsim 10 M_P$.

As usual the cosmological background is given by
\ba
ds^2 = -dt^2  + a(t)^2 d\x^2\,,
\ea
where $a(t)$ is the scale factor.  It is more convenient to change 
the clock from the cosmic time $t$ to the number of $e$-folds $N$
via $d N = H dt$ where $H= \dot a/a$ is the Hubble expansion rate.
Denoting the time when the waterfall phase transition takes place 
by $N_c$, we further define $n\equiv N- N_c$. 
Hence $n<0$ for the period before the phase transition 
and $n>0$ after the transition. We denote the end of the
waterfall transition when $\chi$ has settled down to 
its global minimum by $n=n_f$. With this notation, inflation 
in our model has the three stages: 
(a): $n <0$,  (b): $0 \leq n \leq n_f$ and 
 (c): $n_f < n < N_e-N_c$ where $N_e$ is the time when
the inflation ends. We set $N_e-N_c\sim55$ so that the
waterfall transition falls into the observable range.

We are interested in the limit where the waterfall phase transition is
fairly sharp, corresponding to $n_f \lesssim 1$ so the waterfall
transition and symmetry breaking completion takes place in an $e$-fold
or so. The key ingredient in our analysis is the dynamics of the
waterfall quantum fluctuations during this short period and the curvature
perturbations induced from them. To investigate this we will
use the $\delta N$ 
formalism~\cite{Sasaki:1995aw,Sasaki:1998ug,Wands:2000dp,Lyth-Liddle}.

As mentioned before, we assume that the waterfall field is very heavy
so it stays at $\chi=0$ and $\chi=\chi_{min}$ during the first and 
third stages. It is useful to introduce the dimensionless 
parameters $\alpha$ and $\beta $ by
\ba
\label{alpha-beta}
\alpha \equiv \frac{m^2}{H^2} \simeq \frac{6 M_P^2}{\phi^2} \simeq 
\frac{6 g^2 M_P^2}{M^2}
\,  \quad , \quad
\beta \equiv \frac{M^2}{H^2 } \simeq  \frac{6 M^2 M_P^2}{m^2 \phi^2} 
\simeq \frac{6g^2 M_P^2}{m^2} \, ,
\ea
where in the last approximate equalities in both expressions 
it is assumed that $\phi \simeq \phi_c$.
The assumption that the slow-roll conditions hold during 
the first and third inflationary stages requires $\alpha \ll 1$.
  Demanding that $\beta \gg 1$ for the waterfall field to be heavy, 
we require $g^2 \gg m^2/M_P^2$. On the other hand, from the 
COBE normalization we have $m/M_P \sim 10^{-6}$ so
$g^2 \gg 10^{-12}$. Furthermore, we assume the onset 
of the phase transition and the time of sudden change in the
inflaton mass to occur about 55 $e$-folds before the end
of inflation so that it falls within the CMB observational window.
 In order for inflation to proceed long enough after 
the phase transition as in the standard chaotic inflation we require 
$\phi_c \gtrsim 10 M_P$ so $g^2 \lesssim 10^{-2} M^2/M_P^2$.
 Combining this with 
 $g^2 \gg m^2/M_P^2$ we get $m^2 \ll 10^{-2} M^2$ or $M \gg 10^{-5} M_P$.
 Finally, from the definition of $C$
 we conclude that $g^2/\lambda \ll 10^{-2} C$.
 For example if we take $C\sim10^{-2}$,
we have $g^2/\lambda \ll 10^{-4}$ and
$\lambda=C^{-1}g^2M^2/m^2\sim 10^2g^2M^2/m^2\gg10^4g^2\gg10^{-8}$.

\subsection{Inflaton dynamics} 

As mentioned above during the first and second stage, inflation proceeds
as in chaotic inflation with the potential,
\begin{eqnarray}
V^{-}(\phi) &=& \dfrac{1}{2} \,m^2\phi^2 + \dfrac{M^4}{4 \lambda}.
\end{eqnarray}
During the short second stage $\chi$ grows until the self-interaction term 
$\lambda \chi^4$ becomes important. Then the self-interaction induces
a large mass and $\chi$ settles down to its local minimum.
We denote this time by $n_f$. After that, the waterfall field starts
rolling across the valley determined by $\partial_\chi V(\phi,\chi)=0$.
This gives
\ba 
\chi^2 =\chi_{min}^2\equiv
\dfrac{M^2}{\lambda} -\dfrac{g^2}{\lambda} \phi^2.
\ea
As seen from the above equation the local value of $\chi$ at
the third stage is dictated by the value of the inflaton field.
Accordingly, during the third stage, the inflaton field experiences
an effective potential given by
\begin{eqnarray}
V^{+}_{eff}(\phi) =V(\phi,\chi(\phi))
= \dfrac{1}{2}\,m^2 (1+C)\phi^2 -\dfrac{g^4}{4 \lambda} \phi^4.
\label{Veff}
\end{eqnarray}

Solving the slow-roll equations of motion for $\phi$, we
obtain, for the first and second stages,
\ba
\label{phi1}
-4 M_P^2(N-N_c)=
-4 M_P^2\,n \,
=\phi(n)^2 - \phi_c^2 
\left[ 1 - C \ln \left( \dfrac{\phi}{\phi_c}\right) \right ] \, ,
\ea
whereas for the third stage,
\ba
\label{phi2}
8 M_P^2 (N_e -N) 
&=& -\phi_e^2 +\phi(N)^2 + \dfrac{1+C}{C} \ \phi_c^2 \,
\ln \left[ \dfrac{1-\dfrac{C}{1+C}
 \dfrac{\phi_e^2}{\phi_c^2}}{1- \dfrac{C}{1+C}
\dfrac{\phi^2}{\phi_c^2}}\right]
\\
&=& 
2\phi^2(N)-2\phi_e^2 
 +\dfrac{C}{2} \,\dfrac{\phi^4(N)- \phi_e^4 }{\phi_c^2}+O(C^2)\, .
\label{phi2-app}
\ea

As usual inflation ends at $\phi=\phi_e$ where 
the slow-roll conditions $\epsilon, \eta \ll 1$
are terminated corresponding to $\phi_e = \sqrt{2} M_P$. 
The slow-roll parameters are defined via
\ba
\label{slow}
\epsilon \equiv \frac{M_P^2}{2} \left( \frac{V_\phi}{V} \right)^2 \quad ,
\quad
\eta \equiv M_P^2 \frac{V_{\phi \phi}}{V}  \, .
\ea
In the present case, around the epoch of the phase transition,
the slow-roll parameters are approximately expressed in terms
of the parameter $\alpha$ as
\begin{eqnarray}
\epsilon\simeq\eta\simeq\frac{\alpha}{3}\,.
\label{slparameter}
\end{eqnarray}


\subsection{Waterfall Field Dynamics}

In this section we study the dynamics of the waterfall quantum
fluctuations. One key point in our model, similar to hybrid inflation,
 is that before the waterfall transition $\chi$ is very heavy so that
it firmly stays at its local minimum $\chi=0$
and classically there is no background evolution of the waterfall field.
Following the prescription in \cite{Abolhasani:2010kr} we assume 
that for each horizon-size patch 
one would observe $\delta\chi^2$ as a homogeneous classical background
which varies smoothly over scales larger than the horizon scale.
Thus on a given, sufficiently large scale, say the comoving scale of 
the present Hubble horizon size, one can calculate the mean value
$\langle\delta\chi^2(n) \rangle$ and the fluctuation,
\ba
\label{Delta-def}
\Delta \chi ^2 (n,\bmx)
\equiv \delta \chi ^2 (n,\bmx)- \langle \delta \chi^2(n) \rangle\,,
\ea
where $\langle\delta\chi^2(n) \rangle$
determines the homogeneous background while $\Delta \chi ^2 (n,\bmx)$
gives rise to the curvature perturbations on super-horizon scales.

With this understanding we now look into the dynamics of
the background waterfall field and its quantum fluctuations 
in some details.  The background waterfall dynamics is governed by
\ba
\label{chi-back-eq}
\chi'' + 3 \chi'
+\left( -\beta +g^2 \frac{\phi^2}{H^2} 
 +3\lambda \frac{\chi ^2}{H^2} \right)
\chi =0 \, ,
\ea
where here and below the prime denotes a derivative with respect to $n$.
Neglecting the self-interaction term $\lambda \frac{\chi ^2}{H^2}$
during the second stage when the transition proceeds,
an approximate solution for the background waterfall field
is given by \cite{Abolhasani:2010kr}
\ba
\label{chi-back-n}
\chi (n) \simeq \chi (n=0) \,
\exp \left[\frac{2}{3} \epsilon_{\chi} n^{3/2} \right] \, ,
\ea
where
\ba
\epsilon_{\chi} \simeq \sqrt{\frac{2}{3} \alpha \beta} \, .
\ea

Here it is worth noting the relations among the model parameters.
Our model has four parameters: $M$, $m$, $g$ and $\lambda$.
On the other hand, we have introduced six non-dimensional parameters:
$\alpha$, $\beta$, $\epsilon$, $\eta$, $\epsilon_\chi$
and $C$. Some of these are functions of time, but 
at leading order in the slow-roll approximation, we may 
consider them as constants. Then as a convenient set of 
independent parameters, we may choose $\epsilon$, $C$, $\epsilon_\chi$
and $\lambda$. Then we have
\begin{eqnarray}
&&
\alpha=3\epsilon\,,
\quad
\beta=\frac{\epsilon_\chi^2}{2\epsilon}\,,
\quad
\eta=\epsilon\,,
\quad
g^2=\frac{6\lambda\epsilon^2C}{\epsilon_\chi^2}
\cr
\cr
&&M^2\simeq\frac{12\lambda\epsilon\,C}{\epsilon_\chi^2}M_P^2
\,,\quad
m^2\simeq\frac{72\lambda\epsilon^3C}{\epsilon_\chi^4}M_P^2\,.
\label{pararelations}
\end{eqnarray}

The dynamics of the waterfall field quantum fluctuations
in Fourier space, neglecting the self-interaction term, is governed by
\ba
\label{chi-k-eq0}
\delta \chi_{\bmk}'' + 3 \delta \chi_{\bmk}'
 + \left( \frac{k^2}{a^2 H^2}
 -\beta +g^2 \frac{\phi^2}{H^2} \right)
 \delta\chi_{\bmk}=0 \, ,
\label{mspace}
\ea
where 
\begin{eqnarray}
\delta\chi_{\bmk}
=\int \frac{d^3x}{(2\pi)^{3/2}}\,
\delta\chi({\bmx})e^{-i{\bmk}\cdot{\bmx}}
=a_{\bmk}\chi_k(n)
+a_{-\bmk}^\dag\overline{\chi_k(n)}\,.
\end{eqnarray}
Here $a_{\bmk}$ and $a_{\bmk}^\dag$ are the annihilation
and creation operators, respectively, with respect to
a suitably chosen vacuum and $\chi_k(n)$
is the positive frequency function. Plugging the background 
value of $\phi$ given by Eq.~(\ref{phi1}) into Eq.~(\ref{chi-k-eq0}) 
results in
\ba
\label{chi-k-eq}
\delta \chi_{\bmk}'' + 3 \delta \chi_{\bmk}'
 + \left( \frac{k^2}{k_c^2}e^{-2n}
-\epsilon_{\chi}^2 n \right)
 \delta\chi_{\bmk}=0 \, ,
\ea 
where $k_c$ is the comoving momentum of the mode which exits 
the horizon at the time of waterfall phase transition: $k_c= H a(n=0)$.
As seen from Eq.~(\ref{chi-k-eq}), $\epsilon_{\chi} H^2$ 
measures the effective tachyonic mass of the waterfall field 
when the tachyonic instability develops.
The assumption that the waterfall phase transition is sharp
requires that $\epsilon_{\chi} \gg1$.
The normalization of $\chi_\bmk$ is determined by the canonical
commutation relation, which gives
\begin{eqnarray}
\delta\chi_\bmk\overline{\delta\chi}_\bmk'
-\overline{\delta\chi}_\bmk\delta\chi_\bmk'=\frac{iH^2}{k_c^3e^{3n}}\,.
\label{chi-k-norm}
\end{eqnarray}

For large and negative $n$, Eq.~(\ref{chi-k-eq}) can be
solved by the WKB approximation by setting 
$\delta\chi_\bmk\propto\exp[S_0+S_1+\cdots]$.
Taking account of the normalization condition (\ref{chi-k-norm})
and choosing the standard Minkowski positive frequency
in the limit $n\to-\infty$, the result to first order 
in the WKB approximation is~\cite{Gong:2010zf}
\ba
\label{wkb-app}
\delta \chi_{k}(n)  
= \dfrac{H}{\sqrt{2k_c^3}}
\frac{e^{-3n/2}}{\Bigl((k/k_c)^2e^{-2n}-\epsilon_\chi^2n\Bigr)^{1/4}}
\exp\left[-i \int^n
\left((k/k_c)^2e^{-2n'}-\epsilon_\chi^2n'\right)^{1/2}dn'\right]\,.
\ea

After waterfall phase transition, $n>0$, some (low $k$)
quanta of the waterfall field become tachyonic and 
behave like classical random fields \cite{Abolhasani:2011yp}. 
In particular there are modes which become tachyonic even 
before horizon crossing. 
In order to find the dynamics of the waterfall quantum fluctuations, 
it is convenient to divide the modes into large and small modes
which we denote below by the subscripts $L$ and $S$, respectively.
Large modes are those which exit the horizon before the time
of the phase transition $n=0$ and small modes are those which 
are sub-horizon at $n=0$. 

\begin{figure}
\includegraphics[width =  5in ]{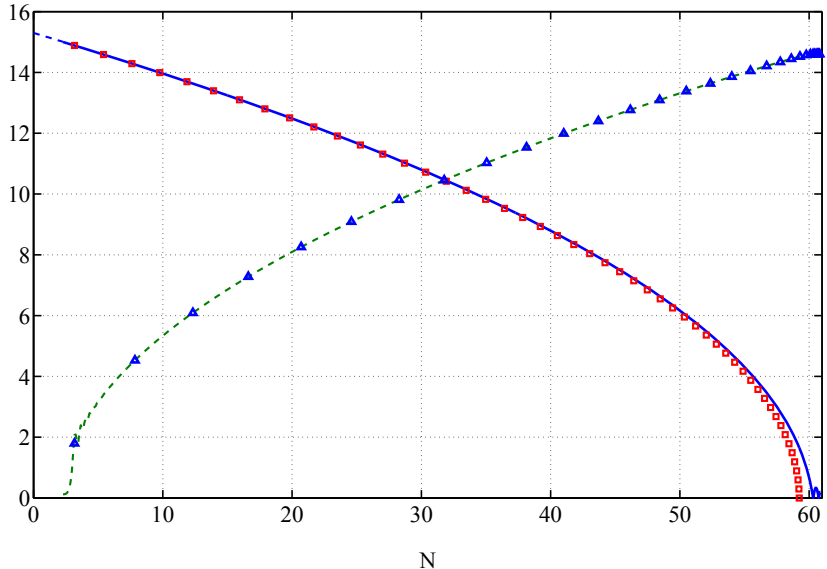}
\caption{
Dynamics of the inflaton and waterfall fields 
(smoothed over a horizon-size patch). The solid blue line 
(the red squares) shows the numerical solution ( analytical 
estimate in Eq. (\ref{phi2}) ) 
for inflaton field . The short dashed blue line for small
 $N$ shows the dynamics of the inflaton field before the 
transition point. As one can see in this figure the inflaton 
dynamics before and after transition are connected smoothly.
 The dashed green line (the blue triangles)
shows our numerical (analytical estimate) results for the 
dynamics of the waterfall field in a smoothing patch. 
In order to visualize this figure better the waterfall field 
amplitude is multiplied by a factor $2 \times 10^3$. 
 The model parameters are $m= 6\times 10^{-6} M_P$, 
$g^2 = 3\times 10^{-8}$, $\lambda =0.14$ and $\phi_c=M/g=15 M_P$,
which give $C=.15$ and $\epsilon_{\chi}=20$.
\label{background}
}

\end{figure}

The large modes cross the horizon sometime before the 
waterfall transition and after horizon crossing their 
profile decays like $\propto e^{-3n/2}$ as seen from
Eq.~(\ref{wkb-app}). At the time of transition, $n=0$,
the WKB approximation fails, but apart from a factor of
order unity, the amplitude at $n=0$ is approximately given by
\ba
\label{delpsi-ls}
|\delta \chi _\bmk^L (n=0)| 
\simeq \dfrac{H}{\sqrt{2\epsilon_\chi k_c^3}} \, .
\ea
After the transition, $n>0$, these modes follow the same
equation as the classical trajectory does.
So by using Eq.~(\ref{chi-back-n}) one has
\ba
\label{delpsi-ls-n}
|\delta \chi _\bmk^L (n>0)| \simeq 
 \dfrac{H}{\sqrt{2\epsilon_\chi k_c^3}}
\exp \left(\frac{2}{3} \epsilon_{\chi}n^{3/2} \right)\,.
\ea

The situation for the small modes is a bit complicated and requires 
careful considerations. As mentioned above the effective tachyonic
 mass of $\delta \chi_\bmk$ is of the order of
$\epsilon_{\chi} H \gg H$.  As a result some modes become
 tachyonic even before horizon crossing.  For modes which 
become tachyonic the spatial gradient term becomes negligible 
compared to the tachyonic mass and the evolution of
$\delta \chi_\bmk$ becomes identical to that of the 
background solution. So for each mode
it is important to find the time when it becomes tachyonic. 
Denoting the time when the mode $k$ becomes tachyonic by $n_t(k)$, 
from Eq.~(\ref{chi-k-eq})  one has
\ba
\label{nt}
n_{t}(k) e^{2 n_{t}(k)} 
= \left(\dfrac{k}{\epsilon_{\chi} k_c} \right)^2 \, .
\ea
Later on we shall call this time the ``classicalization" time. 
The solution of the above algebraic equation are known to be
given by the Lambert W function $W(z)$,
\begin{eqnarray}
n_t(k)=\frac{1}{2}W(z);
\quad z=2\left(\dfrac{k}{\epsilon_{\chi} k_c} \right)^2\,,
\end{eqnarray}
but we only need an approximate solution for $n_t(k)$ since
$n_t(k)\lesssim 1$ for the parameters of our interest.

One can use the WKB solution Eq.~(\ref{wkb-app}) until $n=n_t(k)$,
which gives
\begin{equation}
\delta\chi^S_\bmk(n) = \frac{H}{\sqrt{2k}\,k_c}e^{-n} \,;
\quad n<n_t(k)\,.
\end{equation}
After $n=n_t(k)$ this mode evolves as the classical background 
solution, Eq.~(\ref{chi-back-n}).
By matching the solution with the classical trajectory at 
$n=n_t(k)$, for $n > n_t(k)$ one approximately has
\ba
\label{delpsi-ss-n}
\delta\chi^S_\bmk(n) = \frac{H}{\sqrt{2k}\,k_c}e^{-n_t}
\exp\left[\frac{2}{3} \epsilon_{\chi}(n^{3/2}-n_t^{3/2})\right] \,;
\quad    n> n_t(k)\,.
\ea
Here it is convenient to project $\delta \chi_k^S(n)$ at
time of the onset of waterfall $n=0$ as if all the modes were 
tachyonic at $n>0$, as viewed in \cite{ Abolhasani:2010kr, Gong:2010zf}.
This gives
\ba
\label{del-psi-zero-S}
\delta\chi^S_\bmk(0) = \frac{H}{\sqrt{2k}\,k_c}
\exp\left[-n_t(k)-\frac{2}{3}\, \epsilon_{\chi}n_t^{3/2}(k)\right]\,.
\ea

Because of the exponential growth of the waterfall field
quantum fluctuations  after the transition,
the expectation value $\langle\delta\chi^2\rangle$ 
becomes non-negligible and its rms value soon starts to 
behave as a classical field \cite{Abolhasani:2010kr}
$\chi=\sqrt{\langle\delta\chi^2\rangle}$. 
As a result  $\langle\delta\chi^2\rangle$
is observed as a classical background for an observer 
within each Hubble horizon 
region \cite{Abolhasani:2011yp, Abolhasani:2010kr, Gong:2010zf}.
Therefore it is necessary  to calculate  $\langle \delta \chi^2 \rangle$
given in terms of its power spectrum  ${\cal P}_{\delta \chi}$ as
\ba
\label{chi-exp}
\langle \delta \chi^2 \rangle 
= \int \frac{d^3 k}{(2 \pi)^3} | \delta \chi_\bmk|^2  
= \int \frac{d^3 k}{(2 \pi)^3} P_{\chi}(k)
 = \int \frac{d k}{k} {\cal P}_{\delta \chi}(k) \, ,
\ea 
where the power spectrum is defined by
\ba
\label{power-chi}
\langle \delta \chi_\bmk \delta \chi_\bmq \rangle \equiv 
(2\pi)^3 \delta^3 (\bmk + \bmq) P_{\chi}(k) \,, \quad 
{\cal P}_{\delta \chi} \equiv \frac{k^3}{2 \pi^2} P_{\chi}(k) \, .
\ea
By using Eqs.~(\ref{delpsi-ls-n}) and (\ref{delpsi-ss-n}) one can
 read off the power spectrum of the waterfall quantum fluctuations 
at $n=0$ as
\ba
\label{pr-delpsi}
{\cal P}_{\delta \chi}(k;0) 
=\left\{
\begin{array}{ll}
 \dfrac{H^2}{4 \pi^2\epsilon_\chi}
 \left( \dfrac{k}{k_c}\right)^3\,;
&\quad k <k_c\,,
\\
~\\
\dfrac{H^2\epsilon_{\chi}^2}{4 \pi^2}\,n_t(k) 
\exp\Bigl[-\frac{4}{3} \epsilon_{\chi} n^{3/2}_t(k)\Bigr]
\,;&\quad k > k_c\,.
\end{array}
\right.
\ea
The details of the analysis to calculate $\langle \delta \chi^2 \rangle$
 is given in Appendix~\ref{sec:variance} where it is found that 
 $\langle \delta \chi^2 \rangle$ is dominated
by the small scale modes,
\ba
\label{delchi-var0}
\langle \delta \chi^2 (0) \rangle \simeq 
\langle \delta \chi^2(0) \rangle_S 
 \simeq \dfrac{3 \epsilon_{\chi}^{4/3} H^2}{16 \pi^2} \, .
\ea

As seen from Eq.~(\ref{pr-delpsi}), there is a sharp peak
in the spectrum of waterfall quantum fluctuations.
In order to estimate the width of the peak, we
expand the above spectrum around its peak. 
Solving $\partial {\cal P}_{\delta \chi} / \partial n_t=0$,
the peak position is found as
\ba
\label{k-max}
n_t(k_{max}) = \left(\dfrac{1}{2 \epsilon_{\chi}} \right)^{2/3}\,.
\ea
Expanding the spectrum around this momentum, one finds
\ba
\label{p-gauss-app}
{\cal P}_{\delta \chi}(k;0) \simeq
 \dfrac{H^2\epsilon_{\chi}^2}{4 \pi^2}
   \, \left( \dfrac{1}{2e\, \epsilon_{\chi}}\right)^{2/3}\,
\exp \left[-\dfrac{\left(n_t(k)-n_t(k_{max})\right)^2}{2\sigma^2_{n_t}}
 \right]\,,
\ea
with
\ba
\sigma_{n_t} = \sqrt{\dfrac{2}{3}} 
\left( \dfrac{1}{2\epsilon_{\chi}}\right)^{2/3}
 = \sqrt{\dfrac{2}{3}} n_t(k_{max}) \, .
\ea
But we are interested in the width of spectrum in 
momentum space, $\sigma_{\ast}(k)$.
By using Eq.~(\ref{nt}) it is readily found as
\ba
\label{Delta-n}
\sigma_{\ast}(k)
 &=& \left( 1+ \dfrac{1}{2n_t(k)}\right) \sigma_{n_t}k_{max}
 \\
&\simeq &  \left( \sqrt{\dfrac{1}{6}}
 + {\cal O}(\epsilon_{\chi}^{-2/3}) \right) k_{max}\simeq 0.4 k_{max}\, .
\ea
This indicates that  the width of the waterfall power spectrum 
is independent of the sharpness of the phase transition 
for large $\epsilon_\chi$, which we verified also numerically.


\section{$\delta N$ formalism and curvature perturbations}
\label{sec:deltaN}

In this section, using the $\delta N$ 
formalism~\cite{Sasaki:1995aw}, \cite{Sasaki:1998ug}, \cite{Wands:2000dp} and \cite{Lyth-Liddle},
we calculate the curvature perturbations.
In order to use the $\delta N$ formalism properly we trace back
 the number of $e$-folds
from the end of inflation until the time of horizon crossing
for each  mode.
To avoid confusion we denote the number of $e$-folds 
counted {\it backward}\/ in time from the end of inflation by 
$\calN$, that is, $\calN\equiv N_e-N$.  
Our strategy is to express ${\cal N}$ in terms of the fields
$\phi(n)$ and $\chi^2(n)$ (smoothed on every Hubble patch).

For those modes which exit the horizon after the waterfall
 transition, by using Eq.~(\ref{phi2-app}),
one can easily find the curvature perturbation on comoving slices,
$\calR_c$, as
\ba
\calR_c \,=\, \delta\calN \,= \,  
\left( 1+ \frac{C}{2} \dfrac{\phi^2}{\phi_c^2} \right)
\dfrac{\phi \,\delta \phi}{2 M_P^2} \, ,
\ea
where $\delta\phi$ is to be evaluated on flat hypersurface as usual.
Finding the curvature perturbation for those modes which exit 
the horizon before and during transition needs careful 
calculations. Here, we follow the same step as 
in~\cite{Abolhasani:2011yp} for these modes.

Here it is worth mentioning that 
the duration of the waterfall stage is sensitive to the classical
value of the waterfall field on every smoothing patch.
As discussed before, because of the large tachyonic mass of the
waterfall field there are modes whose spatial gradient can be
neglected and which behave classically already before their 
scales cross the horizon, and hence affect the classical 
trajectory. Noting that the formula $\calR_c= \delta \calN$ 
is valid on scales over which small scale inhomogeneities can be 
smoothed out with a negligible influence on the geometry,
we take the smoothing scale to be slightly larger than the 
comoving scale corresponding to the wavelength of the
last mode which becomes tachyonic.

As we discussed before, during the third inflationary stage
 (after completion of the phase transition) inflation proceeds in
 the form of chaotic inflation with a slight change in the effective 
mass of the inflaton. So the end of inflation is determined by the 
value of $\phi$ alone as
\begin{eqnarray}
\phi=\phi_e\approx \sqrt{2}\,M_P\,.
\end{eqnarray}
From this point up to the time of the end of the waterfall transition,
${\cal N}$ is given by Eq.~(\ref{phi2-app}),
\begin{eqnarray}
{\cal N}=\frac{1}{4M_P^2}
 \left[\phi^2-\phi_e^2
 + \dfrac{C}{4} \dfrac{\phi^4-\phi_e^4}{\phi_c^2} \right]
;\quad {\cal N}\leq {\cal N}_f\,,
\label{afterNp}
\end{eqnarray}
where ${\cal N}_f$ is the value of ${\cal N}$
at the end of the waterfall transition.
Here, for simplicity we consider that at the end of transition
 $\chi$ is very close to its local instantaneous minimum  
so  $\chi^2$ at ${\cal N}={\cal N}_f$ is 
\begin{eqnarray}
\label{psi-min}
\chi^2(n_f)
=\chi_{min}^2(n_f)\, \simeq\,\dfrac{M^2}{\lambda}
 - \dfrac{g^2}{\lambda} \phi_f^2.
\end{eqnarray}
One can obtain ${\cal N}_f$ in terms of the number of $e$-folds 
from the critical epoch $\phi=\phi_c$ to the end of waterfall
 transition, $n_f$, as
\begin{eqnarray}
{\cal N}_f(n_f)=\frac{1}{4M_P^2} 
\left[ \phi_f^2 - \phi_e^2 
+ \dfrac{C}{4 \phi_c^2} \left( \phi_f^4 - \phi_e^4\right) \right ],
\label{Np}
\end{eqnarray}
by which one can readily find
\begin{eqnarray}
4 M_P^2 \delta{\cal N}_f(n_f)= \delta(\phi_f^2)
 \left( 1+\dfrac{C}{2} \dfrac{\phi_f^2}{\phi_c^2} \right) \, . 
\label{del-Np}
\end{eqnarray}
On the other hand using Eq.~(\ref{phi1}) one has
\begin{eqnarray}
-4 M_P^2  \delta n_f \simeq  
\delta \left( \phi_f^2\right) 
\left( 1+\dfrac{C}{2} \dfrac{\phi_c^2}{\phi_f^2} \right) \, .
\label{Np-simp}
\end{eqnarray}
By using the fact that $\phi_f^2 \simeq \phi_c^2 -4M_P^2 n_f $ 
one can find that
\begin{eqnarray}
\delta {\cal N}_f(n_f) \simeq - \, \delta n_f 
   \left( 1- 2 C \epsilon n_f\right) \quad  
\rightarrow \quad 
\dfrac{d{\cal N}_f}{d n_f}  = -1 + 2 C \epsilon \,n_f,
\label{dNf-dnf}
\end{eqnarray}
in which $\epsilon$ denotes the conventional slow-roll parameter
 which is approximately given by
$\epsilon \simeq 2M_P^2/\phi_c^2$.

Now we trace back the evolution to earlier times before the end of 
transition, ${\cal N}>{\cal N}_f$. For this stage, instead
of ${\cal N}$, it is more convenient to use $n$ which is the 
number of $e$-folds from the critical point counted 
{\it forward} in time, i.e., $n=n_f+{\cal N}_f- {\cal N}$. 
Then $\chi^2(n)$ is given by 
\begin{eqnarray}
\label{psi2}
\chi^2(n)=\exp \left[2\left(f(n)-f(n_f) \right) \right]\chi_{min}^2(n_f)\,,
\end{eqnarray}
where $f(n)$ for a sharp phase transition is given by 
\ba
\label{f-n}
f(n) = \frac{2}{3} \epsilon_{\chi} \, n^{3/2} \, .
\ea
During this era, $n$ is expressed in terms of 
$\phi(n)$ as given by Eq.~(\ref{phi1}),
\begin{eqnarray}
-4 M_P^2\,n \,
=\phi(n)^2 - \phi_c^2 
\left[ 1 - C \ln \left( \dfrac{\phi}{\phi_c}\right) \right ] \,.
\label{phi}
\end{eqnarray}
Here we note that $n$ depends on $n_f$ and ${\cal N}$ in
 a non-trivial way,
\begin{eqnarray}
n(n_f,{\cal N})=n_f+ {\cal N}_f(n_f)- {\cal N}\,.
\label{smalln}
\end{eqnarray}
By virtue of the above geometric relation and by using 
Eq.~(\ref{dNf-dnf}) one has
\begin{eqnarray}
\frac{\partial n}{\partial n_f}= 2 C \epsilon \,n_f\,.
\end{eqnarray}

Keeping in mind the above dependence of $n$ on $n_f$ and ${\cal N}$,
let us take the variation of Eqs.~(\ref{psi2}) and (\ref{phi}). 
We obtain
\begin{eqnarray}
\label{dpsi2}
\dfrac{\delta\chi^2(n)}{\langle \delta \chi^2(n) \rangle}
&=& \dfrac{\delta\chi_{min}^2(n_f)}{\chi^2_{min}(n_f)} 
+ 2f'(n)\delta \,n - 2f'(n_f)\delta \,n_f\,,
\\
-2 M_p^2 \delta n 
&\simeq & 
 \phi(n)\delta\phi(n) \left( 1+ \dfrac{C}{2}+C\, \epsilon\, n\right) \,.
\label{dphi}
\end{eqnarray}
On the other hand from Eq.~(\ref{psi-min}), one finds
\begin{eqnarray}
\chi^2_{min}(n_f) = 4 M_P^2 \dfrac{g^2}{\lambda}\,n_f \,,
\end{eqnarray}
where we have used Eq.~(\ref{phi1}) and the fact that 
$\phi_f \lesssim\phi_c$. This results in
\begin{eqnarray}
\dfrac{\delta \chi^2_{min}(n_f)}{\chi^2_{min}(n_f)}
 =  \frac{\delta n_f}{n_f}\,.
\end{eqnarray}

Finally, solving  Eqs.~(\ref{dpsi2}) and (\ref{dphi}) for 
$\delta{\cal N}$, we find
\begin{eqnarray}
\delta {\cal N}=\frac{\delta\chi^2(n)}{\langle \delta\chi^2(n) \rangle}
\dfrac{\partial n}{\partial n_f}
\frac{1}{-2f'(n_f)+n_f^{-1}}
+\frac{\phi\,\delta\phi(n)}{2 M_P^2}
\left[1+\dfrac{C}{2}+C \epsilon n+
\frac{2f'(n)}
{-2f'(n_f)+n_f^{-1}}\frac{\partial n}{\partial n_f}
\right]\,. 
\label{calN1}
\end{eqnarray}
To simplify the above relation we note that 
$ 2n_f f'(n_f) \gg 1$ which is valid in our model with 
a sharp phase transition. Then the above expression reduces to
\begin{eqnarray}
\label{delN-final}
\delta {\cal N}
= -\frac{C \epsilon\,n_f}{f'(n_f)}
\frac{\delta\chi^2(n)}{\langle \delta \chi^2(n)\rangle}
+\left[1+\dfrac{C}{2}+ C \epsilon(n-2n_f) \right]
\frac{\phi \delta\phi}{2 M_P^2}\,.
\end{eqnarray}
Noting that $\delta\chi^2(n)$ and $\langle \chi^2(n)\rangle $ 
have the same $n$-dependence  the final result for the comoving
curvature perturbation ${\cal R}_c = \delta {\cal N}$
can be expressed in terms of $\delta\chi^2(0)$ and 
$\langle \chi^2(0)\rangle $,
\begin{eqnarray}
\label{delN-final-2}
{\cal R}=\delta {\cal N}
= -\frac{C \epsilon\,n_f}{f'(n_f)}
\frac{\delta\chi^2(0)}{\langle \delta \chi^2(0)\rangle}
+\left[1+\dfrac{C}{2}+ C \epsilon(n-2n_f) \right]
\frac{\phi \delta\phi}{2 M_P^2}\,,
\end{eqnarray}
where and below we omit the suffix $c$ from $\calR_c$
and simply denote it by $\calR$ for notational simplicity.

This is our key formula for computing the power spectrum and 
bispectrum. As can be seen from the above expression, the 
curvature perturbation has the conventional inflaton 
contribution up to slow-roll corrections in the inflaton mass 
and the contribution from the waterfall field.
The latter is a dynamical effect which is intrinsic
to our model, in contrast to many other models in which
local features are added simply phenomenologically.

\section{Power Spectrum of Curvature perturbations}

Having found the final curvature perturbations with the $\delta N$
formalism we now calculate the power spectrum.
Our aim is to find an imprint of the sharp waterfall 
transition on the power spectrum.

As clear from Eq.~(\ref{delN-final-2}), the power spectrum 
can be divided into two distinct contributions, the one
from the inflaton and the other from the waterfall field,
\ba
\label{PR}
\nonumber
{\cal P_R} &=& {\cal P}_\calR^{wf}+ {\cal P}_\calR^{\phi}
\\
&=&
\frac{C^2 \epsilon^2 n_f^2}{f'^2(n_f)}~{\cal P}_{\delta\chi^2/\chi^2}
+\left[1+\dfrac{C}{2} +{\cal O}(C \epsilon)\right]^2
\frac{\phi^4}{4 M_P^4}\, {\cal P}_{\delta\phi/\phi}\,.
\ea
Below we evaluate each contribution separately.


\subsection{Contribution of the inflaton to the power spectrum}

Since the inflaton field is light throughout the whole stage of
inflation the amplitude of its quantum fluctuations on flat hypersurface
at the time of horizon crossing is given by the usual formula,
\ba
\delta \phi(k)= \dfrac{H(n_k)}{\sqrt{2k^3}} \,,
\ea
where $H(n_k)$ is the Hubble parameter at the time of horizon crossing.
This gives
\ba
{\cal P}_\calR^{\phi}(k) \simeq \dfrac{1}{4 \pi^2}
\left.\left[1+\dfrac{C}{2}\right]^2
\frac{\phi^2 H^2}{4M_P^4}\right|_{n=n_k}
 \simeq \dfrac{(1+C)}{48 \pi^2} 
\left.\dfrac{\phi^2 \, V^{+}_{eff}(\phi)}{M_P^6}\right|_{n=n_k}\,,
\ea
where $V^{+}_{eff}(\phi)$ is given by Eq.~(\ref{Veff}),
which may be rewritten to first order in $C$ as
\ba
V^{+}_{eff}(\phi) = \dfrac{1}{2} m^2 \left(1+C\right)
\left(1-\frac{C}{2}\frac{\phi^2}{\phi_c^2}\right)\phi^2
+ O(C\, \epsilon) \,.
\ea
Thus we obtain
\ba
\label{inflaton-power}
{\cal P}_\calR^{\phi}(k)= \dfrac{1}{96 \pi^2}
\left(1+C\right)^2 \left(1-\frac{C}{2}\frac{\phi^2}{\phi_c^2}\right)
\left. \frac{ m^2 \phi^4}{M_P^6}\right|_{n=n_k}\,.
\ea
Neglecting the corrections of $O(C)$, 
the above expression reduces to the standard formula,
${\cal P}_\calR^{\phi}(k)=V^3/(12\pi^2V'{}^2M_P^6)|_{n=n_k}$.

\subsection{Contribution of the waterfall to the power spectrum}

Now we calculate the contribution of the waterfall field 
perturbations to the power spectrum. In order to estimate 
this contribution we first note that
\ba
\label{power-chi2}
\Big\langle \left( \delta \chi^2\right)_\bmk 
\left( \delta \chi^2\right)_\bmq  \Big\rangle 
&\equiv& P_{\delta \chi^2}(k) (2 \pi)^3 \delta^3(\bmk+\bmq).\\
\label{power-chi2b}
{\cal P}_{\delta \chi^2/\chi^2}(k)  
&\equiv& \dfrac{1}{\langle  \delta \chi ^2 (0)\rangle^2} \,
 \dfrac{k^3}{2 \pi^2} P_{\delta \chi^2}(k)\,.
\ea
The correlation function of $\delta \chi^2_\bmk$ can be 
calculated using the identity~\cite{Gong:2010zf},
\ba
\label{waterfall-conv}
\Big\langle \left( \delta \chi^2\right)_k
 \left( \delta \chi^2\right)_q  \Big\rangle 
&=& 2 \int \dfrac{d^3q}{(2\pi)^3} 
|\delta \chi_{|\bmk-\bmq|}|^2 |\delta \chi_q|^2
~(2\pi)^3 \delta^3(\bmk+\bmq)\,.
\ea
In Appendix~\ref{sec:app-2} we show that
\ba
\label{p-chi2-approx}
{\cal P}_{\delta \chi^2 }(k) \simeq
 \dfrac{\xi^3 H^2}{4 \pi^2}\, {\cal P}_{\delta \chi}\,,
\ea
where $\xi$ is a numerical factor of order unity.
This means that the power spectrum of $\delta \chi^2$ 
is proportional to the power spectrum of
$\delta \chi$. Plugging this into Eqs.~(\ref{PR}) and
 (\ref{power-chi2b}) and using the explicit form of  $f(n)$ 
given in Eq.~(\ref{f-n}) and $\langle  \delta \chi ^2 (0)\rangle$ 
calculated in Eq.~(\ref{delchi-var}), we find
\ba
\label{waterfall-chi}
{\cal P}_\calR^{wf}(k)
 =  \frac{4 C^2 \epsilon^2 n_f \xi^3}{3 \epsilon_{\chi}^{10/3}}\, 
\dfrac{{\cal P}_{\delta\chi}(k;0)}{\langle \delta \chi^2 (0)\rangle} \,.
\ea
Finally, plugging the waterfall field perturbation 
spectrum~(\ref{pr-delpsi}) to the above, we obtain
\ba
\label{waterfall-power}
{\cal P}_\calR^{wf}(k) 
\simeq\left\{
\begin{array}{ll}
\dfrac{16}{9} C^2 \epsilon^2 n_f \,\epsilon_{\chi}^{-20/3}\, \xi^3 
 \left( \dfrac{k}{k_c}\right)^3\,; 
&\quad k< k_c\,,
\\
~\\
\dfrac{16}{9} C^2 \epsilon^2 n_f \,\epsilon_{\chi}^{-8/3}\, \xi^3
 n_t(k)\exp\left[-\frac{4}{3}\epsilon_{\chi} n_t^{3/2}(k)\right]\,;
&\quad k> k_c\,.
\end{array}
\right.
\ea


\subsection{Total curvature perturbation spectrum}

We now consider the total curvature perturbation
spectrum by adding contributions both from the inflaton field and 
the waterfall field. Since the waterfall contribution is
peaked at $k=k_{max}\gtrsim k_c$, let us first compare the amplitudes
of ${\cal P}_\calR^{wf}(k)$
and ${\cal P}_\calR^{\phi}(k)$ at $k\simeq k_{max}$. 
Using Eq.~(\ref{k-max}) for $k_{max}$ and comparing 
Eqs. (\ref{waterfall-power}) and (\ref{inflaton-power}),
we find
\ba
\label{powers-ratio}
\frac{{\cal P}_\calR^{wf}(k_{max})}{{\cal P}_\calR^{\phi}}
 \simeq 10^{3} C^2 \left(\frac{\epsilon}{10^{-2}}\right)^4
\left(\frac{\epsilon_\chi}{10}\right)^{-10/3} \, ,
\ea
where the approximation $m/M_P \sim 10^{-6}$ has been used 
in order to satisfy the COBE normalization. This result
shows that there can be a prominent peak even for a small $C$,
say $C\sim0.1$.

\begin{figure}
\includegraphics[width =  5in ]{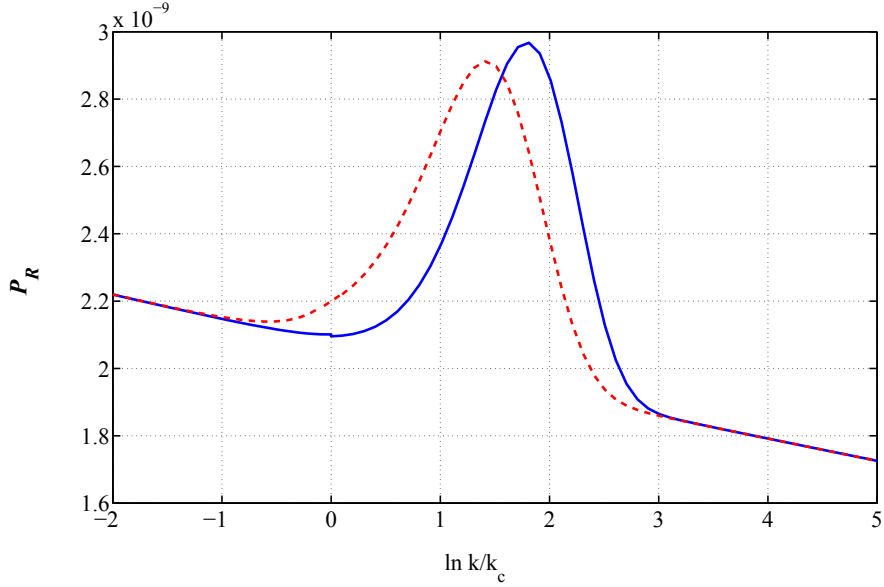}
\caption{
Power spectrum of the curvature perturbation. The dashed red curve shows
an analytical estimate of the total curvature perturbation power 
spectrum by adding both the inflaton and waterfall field contributions, 
${\cal P}_\calR^{\phi}(k) +{\cal P}_\calR^{wf}(k)$, given by
 Eqs.~(\ref{inflaton-power}) and (\ref{waterfall-power}).
The blue solid curve shows the total
 curvature perturbation, ${\cal P_\calR}(k)$, in which the convolution
integral Eq.~(\ref{waterfall-conv}) is numerically calculated.
As one can see in this figure, ${\cal P}_\calR^{wf}(k)$ peaks 
near $k_{max} \sim k_c$ and decays quickly for $k$ not close 
to $k_{max}$. 
The parameters are the same as in Fig.~\ref{background}.}
\label{power-fig1}
\end{figure}

In Fig.~\ref{power-fig1} we plot the total curvature perturbation
power spectrum for the parameters $C=0.15$,
$\epsilon =0.01$ and $\epsilon_\chi =20$. 
 The peak at $k=k_{max} \sim k_c$ in the spectrum is due to 
the waterfall field contribution, ${\cal P}_\calR^{wf}(k)$.
The spectrum away from the peak is dominated by the inflaton 
contribution, ${\cal P}_\calR^{\phi}(k) $.

In passing, it is instructive to estimate $n_f$, the duration
of the phase transition.
As mentioned before, we treat $\langle \delta \chi^2 \rangle $
 as the averaged classical value of the waterfall field on 
each horizon size patch. A good criterion for the completion
of the phase transition is when $\langle \delta \chi^2\rangle $
 reaches the value of its local minimum $\chi_{min}^2(n_f)$ given by
Eq.~(\ref{psi-min}). With the help of Eq.~(\ref{phi}), we find
\begin{eqnarray}
\langle \delta \chi^2(n_f) \rangle\simeq
\frac{4g^2M_P^2}{\lambda}\,n_f\,.
\label{dchi2f}
\end{eqnarray}
In Appendix~\ref{sec:variance} the expectation value
$\langle \delta \chi^2(n) \rangle$ is evaluated as
\ba
\langle \delta \chi^2(n)  \rangle =  \langle \delta \chi^2(0) \rangle 
\exp(\frac{4}{3} \epsilon_\chi n^{3/2}) \simeq
 \frac{3  \epsilon_\chi^{4/3} H^2}{16 \pi^2}
\exp(\frac{4}{3} \epsilon_\chi n^{3/2}) \, .
\ea
Equating this with $\chi_{min}^2(n_f)$, we obtain
an estimate,
\ba
n_f = \Gamma\,\epsilon_\chi^{-2/3} \,;
\quad
\Gamma \simeq \left(\ln 
\left[ \frac{32 \pi^2 \epsilon_\chi^{2/3}}{6\lambda}
\right]\right)^{2/3} \, .
\ea
For our numerical example we find $n_f \simeq 0.5$ so the 
phase transition is fairly sharp. But it is smooth enough
to render the dynamics of the phase transition adiabatic.
Namely, we are free from possible violations of the
adiabaticity of the inflaton vacuum state that may occur
for very sharp transitions as discussed in the 
literature~\cite{Starobinsky:1992ts, Leach:2001zf, Adams:2001vc, Gong:2005jr, Joy:2007na,Chen:2011zf,Chen:2011tu,Romano:2008rr,Barnaby:2009dd,Barnaby:2010ke}.

\section{Bispectrum and non-Gaussianities}
\label{sec:bispectrum}

Now we compute the bispectrum of this model.
Due to the intrinsic non-Gaussian nature of $\delta \chi^2$,
we expect to see large spiky non-Gaussianities when 
${\cal P}_\calR^{wf}(k_{max}) > {\cal P}_\calR^{\phi}(k_{max})$. 

So far in our analysis, we have expanded $\delta N$ up to
 $\delta \chi^2$ as given by Eq.~(\ref{delN-final-2}).
In order to calculate the bispectrum we need to expand $\delta N$ 
up to $\delta \chi^4$. This is done in Appendix~\ref{sec:deltaN-fourth}.
The three point function can be read from Eq.~(\ref{delN-final}) as
\ba
\label{tcp-1}
\nonumber
\langle {\cal R}_{\bmk_1}{\cal R}_{\bmk_2} {\cal R}_{\bmk_3}\rangle 
&\equiv&
B_\calR(\bmk_1,\bmk_2,\bmk_3)(2\pi)^3\delta(\bmk_1+\bmk_2+\bmk_3)
\cr
\cr
&=& (N_{,\chi^2})^3 \big \langle 
\left( \delta \chi^2\right)_{\bmk_1}
\left( \delta \chi^2\right)_{\bmk_2}\left( \delta \chi^2\right)_{\bmk_3}
 \big \rangle 
\nonumber\\
 &+&  \frac{1}{2} (N_{,\chi^2})^2 N_{,\chi^2,\chi^2}
 \Big \langle \left[\left( \Delta \chi^2\right)^2 \right]_{\bmk_1}
 \left( \delta \chi^2\right)_{\bmk_2} \left( \delta \chi^2\right)_{\bmk_3}
+ \mathrm{c.p.} \Big \rangle
\nonumber\\
 &+&  \frac{1}{2} (N_{,\phi})^2 N_{,\phi\,\phi}
 \Big \langle\left(\delta\phi^2\right)_{\bmk_1}
 \delta\phi_{\bmk_2} \delta\phi_{\bmk_3}
+ \mathrm{c.p.} \Big \rangle\,,
\ea
where c.p. represents cyclic permutations,
$(\bmk_1,\bmk_2,\bmk_3)\to(\bmk_2,\bmk_3,\bmk_1)\to(\bmk_3,\bmk_1,\bmk_2)$
and $\Delta \chi^2$ is the fluctuations of $\delta \chi^2(n,\bmx)$ 
on scales larger than the horizon scale, 
as defined in Eq.~(\ref{Delta-def}).

There are two distinct contributions to the three point function. 
The first term in Eq.~(\ref{tcp-1}) is 
 due to the intrinsic non-Gaussianity of $\delta \chi^2$. 
We define the intrinsic bispectrum of $\delta\chi^2$ 
in the standard way by
\begin{eqnarray}
\label{defBpsi}
\Bigl\langle
(\delta\chi^2)_{\bmk_1}(\delta\chi^2)_{\bmk_2}(\delta\chi^2)_{\bmk_3}
\Bigr\rangle
=B_{\delta \chi^2}(k_1,k_2,k_3)
 (2\pi)^3\delta^3(\bmk_1+\bmk_2+\bmk_3) \,.
\end{eqnarray}
The second term in Eq.~(\ref{tcp-1}) is 
due to nonlinear dynamics of the waterfall field,
while the last term is that of the inflaton field
which generically gives a negligible contribution when
the inflaton is slow-rolling. Below we compute the first
and second terms separately.

\subsection{Dynamically generated bispectrum}
\label{subsec:dynamicalB}

Let us first concentrate on the second term,
\ba
\label{dynB}
\langle {\cal R}_{\bmk_1}{\cal R}_{\bmk_2} {\cal R}_{\bmk_3}\rangle_{(2)} 
\equiv
\frac{1}{2} (N_{,\chi^2})^2 N_{,\chi^2,\chi^2}
 \Big \langle \left[\left( \Delta \chi^2\right)^2 \right]_{\bmk_1}
 \left( \delta \chi^2\right)_{\bmk_2} \left( \delta \chi^2\right)_{\bmk_3}
+ \mathrm{c.p.} \Big \rangle\,.
\ea
Here we have intentionally avoided to call the above the dynamically 
generated bispectrum, because we shall see that it also includes
some contribution from the intrinsic bispectrum of $\delta\chi^2$. 

By noting that 
\ba
\left[\left( \Delta \chi^2\right)^2 \right]_{\bmk}
= \left(\delta \chi^4 \right)_{\bmk}- 2 
\langle \delta \chi^2 \rangle \left( \delta \chi^2\right)_{\bmk}\,,
\ea
one has
\ba
&&\left\langle \left[\left( \Delta \chi^2 \right)^2 \right]_{\bmk_1} 
\left( \delta \chi^2 \right)_{\bmk_2} \left( \delta \chi^2\right)_{\bmk_3}
+ \mathrm{c. p.} \right\rangle 
\nonumber\\
&&\quad= -2 \times3 \times
 \langle\delta \chi^2 \rangle \big \langle
 \left( \delta \chi^2\right)_{\bmk_1}\left( \delta \chi^2\right)_{\bmk_2}
\left( \delta \chi^2\right)_{\bmk_3} \big \rangle  
+\Biggl[ \big \langle \left( \delta \chi^4\right)_{\bmk_1}
\left( \delta \chi^2\right)_{\bmk_2}\left( \delta \chi^2\right)_{\bmk_3}
 \big \rangle 
+\mathrm{c.p.}\Biggr]
\nonumber\\
&&\quad
= -6\langle\delta \chi^2 \rangle B_{\delta \chi^2}(k_1,k_2,k_3)
 \,(2\pi)^3\delta^3(\bmk_1+\bmk_2+\bmk_3) 
+\Biggl[ \bigl\langle ( \delta \chi^4)_{\bmk_1}
(\delta\chi^2)_{\bmk_2}(\delta \chi^2)_{\bmk_3}
 \bigr\rangle 
+\mathrm{c.p.}\Biggr]\,.
\label{dynamic3pt}
\ea
The first term on the right hand side above, proportional to 
$B_{\delta\chi^2}$, is of the same form as the first term in
Eq.~(\ref{tcp-1}), which we evaluate later.
Here we first focus on the other terms in the square brackets.

We note that if $\delta \chi^2$ were Gaussian, we could
express these terms in terms of the product of two 
point correlation functions. But since this is not the case
 for $\delta \chi^2$, there is also a contribution proportional
 to the three point correlation function of $\delta\chi^2$,
as we shall see below.
 
By expanding $\delta \chi^4$ one has
\ba
\left\langle \left( \delta \chi^4 \right)_{\bmk_1}
 \left( \delta \chi^2 \right)_{\bmk_2}
 \left( \delta \chi^2 \right)_{\bmk_3}\right\rangle
  =\int \widetilde {d^3 q} \, \left\langle \delpsi_{\bmk_1-\bmq} 
\delpsi_{\bmq} \delpsi_{\bmk_2} \delpsi_{\bmk_3} \right\rangle\,,
\ea
in which we have introduced
$\widetilde{d^3 q} = {d^3 q}/(2\pi)^3$ to simplify the notation.
To calculate the r.h.s of this equation we should first 
classify possible contractions. Since we are not interested in
tad-pole type graphs but only in irreducible graphs, 
there are some restrictions on non-trivial contractions. 
First, contractions between the terms within any of
 $(\delta\chi^2)_{\bmp}$ themselves are not allowed. 
Second, contractions should not be closed only within the terms 
in $(\delta\chi^4)_{\bmk_1}$, corresponding to the first two terms 
in the r.h.s. of the equation.
Then one finds there are two different classes of contractions.
The first class is in which there is one contraction between 
a pair of terms in $(\delta\chi^4)_{\bmk_1}$.
The second class is in which there is no contraction between any
pair of terms in $(\delta\chi^4)_{\bmk_1}$ themselves.

Let us consider the first class and count the number of possible
contractions.
There are 4 choices to choose a pair in $(\delta\chi^4)_{\bmk_1}$.
Then one of the remaining two $\delta\chi$ has 4 choices to contract
with one of $\delta \chi$ in $(\delta\chi^2)_{\bmk_2}$ and 
$(\delta\chi^2)_{\bmk_3}$, and the last $\delta\chi$ in 
$(\delta\chi^4)_{\bmk_1}$ has 2 choices to contract the remaining 
terms in $(\delta\chi^2)_{\bmk_2}$ and $(\delta\chi^2)_{\bmk_3}$.
Thus there are in total $4\times4\times2=32$ possible contractions
in the first class. They all give the same result.
So let us calculate one of them:
\begin{eqnarray}
&&\int \widetilde{d^3 q} \prod_{i} \widetilde{d^3 p_i}
\left\langle \wick{2121}{\Bigl( <1\delta \chi _{\bmk_1- \bmp_1 -\bmq}
\delta <2 \chi_{\bmp_1}\big) \big( \delta >1 \chi_{\bmq- \bmp_2}
\delta <3 \chi_{\bmp_2} \big) \big( \delta >2 \chi_{\bmk_2- \bmp_3}
\delta <4 \chi_{\bmp_3} \big) \big( \delta >3 \chi_{\bmk_3- \bmp_4}
\delta >4\chi_{\bmp_4}\Bigr) }\right\rangle 
\nonumber\\
&&=\int \widetilde{d^3 q} \prod_{i}d^3 p_i\,
 \vert \delta \chi _{\vert \bmq - \bmp_{2} \vert}\vert^2 
\vert \delta \chi _{p_1}\vert^2 \vert \delta \chi _{p_2}\vert^2
\vert \delta \chi _{p_{3}} \vert^2 \times \mbox{$\delta$-factor}\,,
\end{eqnarray}
where 
\begin{eqnarray}
\mbox{$\delta$-factor}
=\delta^3(\bmk_{1}-\bmp_{1}-\bmp_{2})
\delta^3(\bmk_{2}+\bmp_{1}-\bmp_{3})
\delta^3(\bmk_{3}+\bmp_{2}-\bmp_{4})\delta^3(\bmp_{3}+\bmp_{4})\,.
\end{eqnarray}
Performing first the integrals over $\bmp_{2}$, $\bmp_3$ and $\bmp_4$,
and then over $\bmq$, the above reduces to
\ba
&&\int \widetilde{d^3 q}\,\int d^3 p_1\,
\vert \delta \chi _{\vert \bmq + \bmp_1-\bmk_1\vert}\vert^2
\vert \delta \chi _{p_1}\vert^2 
\vert \delta \chi _{\vert \bmk_{2}+\bmp_{1} \vert}\vert^2
\vert \delta \chi _{\vert \bmk_{1}-\bmp_{1} \vert} \vert^2 
\delta^3(\bmk_{1}+\bmk_{2}+\bmk_{3})
\nonumber\\
&&\quad=\langle \delta \chi^2 \rangle
\int d^3 p_1\,
\vert \delta \chi _{p_1}\vert^2 
\vert \delta \chi _{\vert \bmk_{2}+\bmp_{1} \vert}\vert^2
\vert \delta \chi _{\vert \bmk_{1}-\bmp_{1} \vert} \vert^2 
 \,\delta^3(\bmk_{1}+\bmk_{2}+\bmk_{3})
\nonumber\\
&&\quad
=\frac{1}{8}\langle \delta \chi^2 \rangle
\big \langle \left( \delta \chi^2 \right)_{\bmk_1} 
\left( \delta \chi^2 \right)_{\bmk_2}
 \left( \delta \chi^2 \right)_{\bmk_3} \big \rangle
\,,
\ea
where we have used the fact that \cite{Gong:2010zf} 
\begin{eqnarray}
\big \langle \left( \delta \chi^2 \right)_{\bmk_1} 
\left( \delta \chi^2 \right)_{\bmk_2}
 \left( \delta \chi^2 \right)_{\bmk_3} \big \rangle
=8\int d^3 p_1\,
\vert \delta \chi _{p_1}\vert^2 
\vert \delta \chi _{\vert \bmk_{2}+\bmp_{1} \vert}\vert^2
\vert \delta \chi _{\vert \bmk_{1}-\bmp_{1} \vert} \vert^2 
 \delta^3(\bmk_{1}+\bmk_{2}+\bmk_{3})\,.
\end{eqnarray}
Thus we obtain
\ba
\nonumber
&&\int \widetilde{d^3 q} \prod_{i} \widetilde{d^3 p_i} 
\langle\wick{2121}{\big( <1\delta \chi _{\bmk_1- \bmp_1 -\bmq} 
 \delta <2 \chi_{\bmp_1}\big) \big( \delta >1 \chi  _{\bmq- \bmp_2} 
 \delta <3 \chi_{\bmp_2} \big) \big( \delta >2 \chi _{\bmk_2- \bmp_3} 
 \delta <4 \chi_{\bmp_3} \big) \big( \delta >3 \chi _{\bmk_3- \bmp_4} 
 \delta >4\chi_{\bmp_4}\big) }\rangle 
\\
&&\qquad
=\frac{1}{8} \langle \delta \chi^2 \rangle
 B_{\delta \chi^2}(k_1,k_2,k_3) \,(2\pi)^3 
\delta^3(\bmk_1+\bmk_2+\bmk_3)\,,
\ea
where the bispectrum $B_{\delta \chi^2}(k_1,k_2,k_3)$
is defined in Eq.~(\ref{defBpsi}).
Since there are $32\times3$ of the same terms, the 
contribution from this class amounts to
\begin{eqnarray}
\label{1sttotal}
\Biggl[ \left\langle ( \delta \chi^4)_{\bmk_1}
(\delta\chi^2)_{\bmk_2}(\delta \chi^2)_{\bmk_3}
 \right\rangle 
+\mathrm{c.p.}\Biggr]_{\mbox{1st}}
&=&32\times3\times\frac{1}{8} \langle \delta \chi^2 \rangle
 B_{\delta \chi^2}(k_1,k_2,k_3) \,(2\pi)^3 
\delta^3(\bmk_1+\bmk_2+\bmk_3)
\nonumber\\
&=&12 \,\langle \delta \chi^2 \rangle
 B_{\delta \chi^2}(k_1,k_2,k_3) \, (2\pi)^3 
\delta^3(\bmk_1+\bmk_2+\bmk_3)\,.
\end{eqnarray}


Let us now turn to the second class, in which every
$\delta\chi$ in $(\delta\chi^4)_{\bmk_1}$ is contracted with
one of $\delta\chi$ in either $(\delta\chi^2)_{\bmk_2}$ or
$(\delta\chi^2)_{\bmk_3}$.
Hence, there are $4\times3\times2=24$ possible contractions.
Again they all give the same result. One of them
is given by
\ba
&&\int \widetilde{d^3 q} \prod_{i} \widetilde{d^3 p_i }
\left\langle \Wwick{12}{
\big(<1\delta \chi _{\bmk_1- \bmp_1 -\bmq} \delta <2 \chi_{\bmp_1}\big) 
\big( \delta  \chi  _{\bmq- \bmp_2}  \delta  \chi_{\bmp_2} \big) 
\big( \delta >1 \chi _{\bmk_2- \bmp_3}  \delta  \chi_{\bmp_3} \big) 
\big( \delta >2 \chi _{\bmk_3- \bmp_4}  \delta \chi_{\bmp_4}\big) }
{12}{ \big( \delta \chi _{\bmk_1- \bmp_1 -\bmq}  \delta \chi_{\bmp_1}\big)
 \big( \delta <3 \chi  _{\bmq- \bmp_2}  \delta <4 \chi_{\bmp_2} \big)
 \big( \delta  \chi _{\bmk_2- \bmp_3}  \delta >3 \chi_{\bmp_3} \big) 
\big( \delta  \chi _{\bmk_3- \bmp_4}  \delta >4\chi_{\bmp_4}\big)
 }\right\rangle
\nonumber
\\
&&=
\int \widetilde{d^3 q} \prod_{i} d^3 p_i\,
\vert \delta \chi _{\vert \bmk_{2} - \bmp_{3} \vert}\vert^2
 \vert \delta \chi _{p_1}\vert^2 \vert \delta \chi _{p_2}\vert^2
\vert \delta \chi _{p_{3}} \vert^2 \times \mbox{$\delta$-factor}\,,
\ea
where
\begin{eqnarray}
\mbox{$\delta$-factor}=
\delta^3(\bmk_{1}+\bmk_{2}-\bmp_{1}-\bmp_{3}-\bmq)
\delta^3(\bmk_{3}+\bmp_{1}-\bmp_{4})
\delta^3(\bmq-\bmp_{2}+\bmp_{3})\delta^3(\bmp_{2}+\bmp_{4})\,.
\end{eqnarray}
Performing first the integrals over $\bmp_{2}$, $\bmp_3$ and $\bmp_4$,
and then over $\bmq$ gives
\ba
&&\int \widetilde{d^3p_1}\,
 \vert \delta\chi _{p_1}\vert ^2
 \vert \delta\chi _{|\bmk_{3}+\bmp_{1}|}\vert ^2 \,
\int d^3q \vert \delta\chi_{|\bmq + \bmp_{1}-\bmk_{1}|}\vert ^2
 \vert \delta\chi _{|\bmq+\bmp_{1}+\bmk_{3}|}\vert ^2
\,\delta ^3(\bmk_{1}+\bmk_{2}+\bmk_{3})
\nonumber\\
&&\quad= 
\frac{1}{4} P_{\delta\chi^2}(k_2)\, P_{\delta\chi^2}(k_3)
 \,(2\pi)^{3}\delta^3(\bmk_{1}+\bmk_{2}+\bmk_{3})\,,
\ea
where we have used the definition of $P_{\delta\chi^2}$,
 Eq.~(\ref{power-chi2}).
Since there are 24 of them, plus cyclic permutations, the contribution
from this class is in total,
\begin{eqnarray}
\label{2ndtotal}
&&\Biggl[ \left\langle ( \delta \chi^4)_{\bmk_1}
(\delta\chi^2)_{\bmk_2}(\delta \chi^2)_{\bmk_3}
 \right\rangle 
+\mbox{c.p.}\Biggr]_{\mbox{2nd}}
\nonumber\\
&&\qquad= 
24\times\frac{1}{4} \Bigl[P_{\delta\chi^2}(k_2)\, P_{\delta\chi^2}(k_3)
 +\mbox{c.p.}\Bigr]\,(2\pi)^{3}\delta^3(\bmk_{1}+\bmk_{2}+\bmk_{3})
\nonumber\\
&&\qquad=6\Bigl[P_{\delta\chi^2}(k_1)\, P_{\delta\chi^2}(k_2)
+\mbox{c.p.}\Bigr]
 \,(2\pi)^{3}\delta^3(\bmk_{1}+\bmk_{2}+\bmk_{3})\,.
\end{eqnarray}

Adding up all the contributions given by
Eqs.~(\ref{1sttotal}) and (\ref{2ndtotal})
together with the first term in (\ref{dynamic3pt}),
we obtain
\ba
&&\Bigl\langle \left[( \Delta\chi^2)^2 \right]_{\bmk_1} 
( \delta \chi^2)_{\bmk_2}(\delta \chi^2)_{\bmk_3}
+ \mbox{c.p.} \Bigr\rangle
\nonumber\\
&&=6 \left( \langle\delta\chi^2\rangle 
B_{ \delta \chi^2 } \left(k_1,k_2,k_3 \right)
+\Bigl[P_{\delta\chi^2}(k_1)\,P_{\delta\chi^2}(k_2) 
+\mbox{c.p.}\Bigr] \right)\,
(2\pi)^{3}\delta^3(\bmk_{1}+\bmk_{2}+\bmk_{3}) \, .
\ea
Substituting the above in Eq.~(\ref{dynB}), the three point
 correlation function of the curvature perturbation
can be represented in terms of the power spectrum and bispectrum of 
$\delta \chi^2$ as
\ba
\label{tcp-2}
\langle{\cal R}_{\bmk_1}{\cal R}_{\bmk_2}{\cal R}_{\bmk_3}\rangle_{(2)}
&=&
\Biggl[3(N_{,\chi^2})^2N_{,\chi^2\chi^2}\langle\delta\chi^2\rangle
 B_{\delta\chi^2}(k_1,k_2 ,k_3)
\nonumber\\
&&\quad
+3(N_{,\chi^2})^2N_{,\chi^2 \chi^2}
\Bigl( P_{\delta\chi^2}(k_1)\,P_{\delta\chi^2}(k_2) +\mbox{c.p.}\Bigr) 
\Biggr]\, (2 \pi)^3 \delta^3 \left(\bmk_1+\bmk_2+\bmk_3 \right)\,.
\ea
As it is clear in the above result, while the second term
is an ordinary nonlinear interaction term with the vertex
proportional to $N_{,\chi^2\chi^2}$, which can be easily
evaluated, the first term is due to the intrinsic non-Gaussianity of 
the $\delta\chi^2$ field which needs to be computed. 

The above equation can be further simplified by using Eq.~(\ref{dN-dpsi})
and noting that $P_{\cal R}^{wf} = N^2_{,\chi^2}P_{\delta\chi^2}$, 
\ba
\langle{\cal R}_{\bmk_1}{\cal R}_{\bmk_2}{\cal R}_{\bmk_3}\rangle_{(2)}
=-3 (N_{,\chi^2})^3 B_{\delta\chi^2}(k_1,k_2 ,k_3)
(2 \pi)^3 \delta^3 \left(\bmk_1+\bmk_2+\bmk_3 \right)
+\langle{\cal R}_{\bmk_1}{\cal R}_{\bmk_2}{\cal R}_{\bmk_3}\rangle_{dyn}
\label{Rbispec2}\,,
\ea
where we have defined the dynamically generated bispectrum
of the curvature perturbation by
\ba
\langle{\cal R}_{\bmk_1}{\cal R}_{\bmk_2}{\cal R}_{\bmk_3}\rangle_{dyn}
&=&B_\calR^{dyn}(\bmk_1,\bmk_2,\bmk_3)
(2 \pi)^3 \delta^3 \left(\bmk_1+\bmk_2+\bmk_3 \right)
\cr\cr
&\equiv&
3\frac{N_{,\chi^2\chi^2}}{(N_{,\chi^2})^2}
\Bigl(P^{wf}_{\cal R}(k_1)P^{wf}_{\cal R}(k_2) +\mbox{c.p.}\Bigr)
(2 \pi)^3 \delta^3 \left(\bmk_1+\bmk_2+\bmk_3 \right)
\,.
\label{Bdyndef}
\ea
%

\subsection{Bispectrum from intrinsic non-Gaussianity}

Now we evaluate the bispectrum from the intrinsic
non-Gaussianity of $\delta\chi^2$. From Eqs.~(\ref{tcp-1}) and
(\ref{Rbispec2}), we have
\begin{eqnarray}
\langle{\cal R}_{\bmk_1}{\cal R}_{\bmk_2}{\cal R}_{\bmk_3}\rangle_{int}
&=&B_\calR^{int}(\bmk_1,\bmk_2,\bmk_3)(2\pi)^3\delta(\bmk_1+\bmk_2+\bmk_3)
\cr
\cr
&\equiv&
-2(N_{,\chi^2})^3
B_{\delta\chi^2}(\bmk_1,\bmk_2,\bmk_3)
(2 \pi)^3 \delta^3 \left(\bmk_1+\bmk_2+\bmk_3 \right)
\,.
\label{Bint}
\end{eqnarray}
Following \cite{Gong:2010zf}, we obtain an expression
for the bispectrum of $\delta \chi^2$ as
\ba
\label{corr-3}
B_{\delta \chi^2}(k_1,k_2,k_3)
=8 \int\widetilde{d^3q}\,
\vert \delta \chi_q\vert^2 
\vert \delta \chi_{\vert \bmk_1- \bmq\vert}\vert^2
\vert \delta \chi_{\vert \bmk_2+ \bmq\vert}\vert^2 \, .
\ea
It is hard to calculate the above integral in general, but 
it may be evaluated in the squeezed limit, 
say $k_3\ll k_1=k_2\equiv k$. 
In this limit, the above reduces to 
\ba
\label{conv-squeezed}
B^{\rm sq}_{\delta\chi^2}(k)\equiv
B_{\delta \chi^2}(k,k,0)\simeq 8 \int \widetilde{d^3 q}\,
\vert \delta \chi_q\vert^2 
\vert \delta \chi_{\vert \bmk-\bmq\vert}\vert^4 \, .
\ea
In Appendix~\ref{sec:app-2}, this integral is evaluated,
and the result (given by Eq.~(\ref{B-squeezed})) is
\ba
B^{\rm sq}_{\delta \chi^2}(k) \simeq 
\dfrac{\xi'^3}{k_c^3} \dfrac{H^4}{2 \pi^2} P_{\delta \chi} (k) 
\simeq \left(\dfrac{\xi'}{\xi} \right)^3 
\dfrac{2 H^2}{k_c^3} P_{\delta \chi^2} (k)\,,
\ea
where the second step follows from Eq.~(\ref{p-chi2-app}) or
(\ref{p-chi2-approx}), and $\xi$ and $\xi'$ are constants of order unity.

Another limiting case of interest is when the magnitudes
of all the momenta are equal to each other, $k_1=k_2=k_3$,
the so-called equilateral limit. Although we have no clue whatsoever
to evaluate the bispectrum in this limit, it is plausible that
the amplitude is at most of the same order as that in the squeezed 
limit, if not much smaller. So let us set
\begin{eqnarray}
\label{B-int}
B^{\rm eq}_{\delta \chi^2}(k) 
=\xi^{\rm eq}\dfrac{2 H^2}{k_c^3} P_{\delta \chi^2} (k)\,,
\end{eqnarray}
where $\xi^{\rm eq}$ is a non-dimensional factor supposedly
of order unity.

\subsection{Total bispectrum and $f_{NL}$ parameter}

It is customary to express the bispectrum of the curvature 
perturbation in terms of the non-Gaussianity parameter 
$f_{NL}$~\cite{Komatsu:2001rj}.
The standard definition of the non-Gaussianity parameter in Fourier
space is \cite{Lyth-Liddle}
\ba
\frac{6}{5}f_{NL} (k_1,k_2,k_3) 
= \frac{ B_{\cal R}(k_1,k_2,k_3)}
{\left[P_{\cal R}(k_1)P_{\cal R}(k_2)+ \mbox{c.p.}\right]}\,.
\ea
In our case we may decompose it as
\ba
f_{NL} = f^{int}_{NL}+f^{dyn}_{NL}+f^{\phi}_{NL}\,,
\ea
where
\ba
\label{fnl-int}
\frac{6}{5}f^{int}_{NL}
 &=&\frac{ B_{\cal R}^{int}(k_1,k_2,k_3)}
{\left[P_{\cal R}(k_1)P_{\cal R}(k_2)+ \mbox{c.p.}\right]}
=-2 \left( N_{,\chi^2} \right)^3
\frac{B_{\delta\chi^2}(k_1,k_2,k_3)}
{\left[P_{\cal R}(k_1)P_{\cal R}(k_2)+ \mbox{c.p.}\right]}\,,
\\
\cr
\label{fnl-dyn}
\frac{6}{5}f^{dyn}_{NL}
&=&\frac{ B_{\cal R}^{dyn}(k_1,k_2,k_3)}
{\left[P_{\cal R}(k_1)P_{\cal R}(k_2)+ \mbox{c.p.}\right]}
=3 \frac{N_{,\chi^2\chi^2}}{ (N_{,\chi^2})^2}\,
\dfrac{\Bigl[P^{wf}_{\cal R}(k_1)
P^{wf}_{\cal R}(k_2) +\mbox{c.p.}\Bigr]}
{\Bigl[P_{\cal R}(k_1)P_{\cal R}(k_2) +\mbox{c.p.}\Bigr]}
\\
\cr
\frac{6}{5}f^{\phi}_{NL}
&=&\frac{ B_{\cal R}^{\phi}(k_1,k_2,k_3)}
{\left[P_{\cal R}(k_1)P_{\cal R}(k_2)+ \mbox{c.p.}\right]}
=\dfrac{1}{2} \frac{N_{,\phi\,\phi}}{(N_{,\phi})^2}
\dfrac{\Bigl[P^{\phi}_{\cal R}(k_1)
P^{\phi}_{\cal R}(k_2) +\mbox{c.p.}\Bigr]}
{\Bigl[P_{\cal R}(k_1)P_{\cal R}(k_2) +\mbox{c.p.}\Bigr]}\,.
\ea
As noted before, the inflaton contribution $f_{NL}^\phi$
is known to be at most of the order of the slow-roll parameters,
$f_{NL}^\phi=O(\epsilon,\eta)$ \cite{Maldacena:2002vr}
hence can be safely neglected.
So we focus on the contributions from the waterfall fields.

First let us consider $f_{NL}^{dyn}$. 
As clear from its form, it is non-negligible only when the
amplitude of $P_\calR^{wf}(k)$ is comparable to or greater 
than that of $P_\calR^{\phi}(k)$, and this may happen
only at and around the peak of the spectrum $k=k_{max}$. 
Then it is easy to see that $f_{NL}^{dyn}$ is non-negligible
only when all of $k_1$, $k_2$ and $k_3$ are approximately equal to
$k_{max}$. Thus setting $k_1=k_2=k_3=k$
and assuming $P_\calR^{wf}(k)$ dominates the spectrum,
we find
\ba
\label{ng-dyn}
\frac{6}{5}f^{dyn}_{NL}(k=k_{max})
 \simeq 3 \frac{N_{,\chi^2\chi^2}}{\left( N_{,\chi^2}\right)^2}
=-\dfrac{3\epsilon_{\chi}}{C\,\epsilon\,n_f^{1/2}} \,,
\ea
where we have used Eq.~(\ref{dN-dpsi}) in Appendix~\ref{sec:deltaN-fourth}
to eliminate $N_{,\chi^2}$ and $N_{,\chi^2\chi^2}$
from the intermediate expression. With $\epsilon_\chi \gg 1, C\ll1$ 
and $n_f \sim1$, 
we see that $f^{dyn}_{NL}$ can become very large, centered around 
$k = k_{max}$. For example,
for the parameters used in our numerical analysis, we have
\ba
\frac{6}{5}f_{NL}^{dyn}(k_{max}) \simeq 7 \times 10^4 \, .
\ea

Next we consider $f_{NL}^{int}$. As clear from the form of 
$B_{\delta\chi^2}$ in Eq.~(\ref{Bint}), or its explicit
evaluation, it remains finite in the squeezed limit,
while the denominator of $f_{NL}^{int}$ diverges
since $P_\calR^{\phi}$ is approximately proportional to $k^{-3}$.
Therefore, $f_{NL}^{int}$ is completely negligible in the
squeezed limit. 
In contrast, the denominator is finite and of order $k^{-6}$
in the equilateral limit. Hence, using Eq. (\ref{B-int}), we obtain an estimate
\ba
\frac{6}{5}f^{int}_{NL}(k)
\simeq \xi^{\rm eq}N_{,\chi^2}\,\dfrac{2H^2}{3k_c^3}
\frac{P^{wf}_{\cal R} (k)}{P_{\cal R}^2(k)}
=\xi^{\rm eq}N_{,\chi^2}\,\dfrac{2H^2}{3}\frac{k^3}{k_c^3}
\frac{{\cal P}^{wf}_{\cal R} (k)}{{\cal P}_{\cal R}^2(k)}
\,.
\ea
Again, this contribution to the non-Gaussianity is  
exponentially negligible except at around the peak of the 
waterfall field spectrum. Hence assuming 
${\cal P}_\calR^{wf}\simeq{\cal P}_{\cal R}$, and manipulating
with the help of Eqs.~(\ref{waterfall-chi}), (\ref{waterfall-power}) and (\ref{dN-dpsi}),
the above estimate gives
\ba
\label{ng-int}
f^{int}_{NL}(k_{max}) \sim
\xi^{\rm eq} N_{,\chi^2}\,\dfrac{H^2 k^3}{k_c^3}
\frac{1}{{\cal P}_{\cal R}(k)} \sim \xi^{\rm eq} \epsilon_\chi^2 f^{dyn}_{NL}(k_{max}) 
\,,
\ea
Depending on the value of the parameter $\xi^{\rm eq}$ of which
we have no quantitative estimate, the intrinsic non-Gaussianity
can be larger than the dynamical non-Gaussianity.

In any case, we conclude that the total non-Gaussianity parameter 
$f_{NL}$ is at least as big as
\ba
f_{NL}(k_{max}) = f^{int}_{NL} + f^{dyn}_{NL}
\sim  \dfrac{\epsilon_{\chi}}{C \epsilon \sqrt{n_f}} \, .
\ea
The width of this sharp feature in the bispectrum is
the same as that of the spectrum, Eq.~(\ref{Delta-n}),
ie, $\sigma(k)\sim 0.4k_{max}$.


\section{Conclusion and Discussions}
\label{sec:conclusion}

In this paper we presented a dynamical mechanism to generate 
a sharp feature during inflation. The key role is played by the 
waterfall field $\chi$ which becomes tachyonic during inflation.
We work in a region of the parameter space where the waterfall 
transition is fairly sharp, i.e., the duration of the transition
is about one $e$-fold or so. Because of the coupling 
$g^2 \phi^2 \chi^2$ the phase transition induces a sharp but small
change in the inflaton mass. 
In much of previous works in which sharp changes in the
inflaton mass were studied, the focus was on the dynamics
of the inflaton itself. In contrast, in our model a local feature 
is induced by non-trivial interactions between the waterfall
and inflaton fields. In particular, the waterfall quantum 
fluctuations played the key role in determining the local feature.

Before the phase transition $\chi$ is very massive, so
it has no classical evolution. It stays at its local minimum at $\chi=0$.
When the inflation passes a critical value, $\chi$ becomes tachyonic
and the waterfall transition commences. 
Then the squared fluctuation, $\langle \delta \chi^2 \rangle$,
starts to grow exponentially and the effective classical trajectory 
is determined by $\langle \delta \chi^2 \rangle$ averaged
over each Hubble horizon patch. The fluctuations 
from one horizon region to another,
$\Delta \chi^2 = \delta \chi^2 - \langle \delta \chi^2 \rangle $,
determines the fluctuation around the classical trajectory.

We calculated the power spectrum of $\Delta \chi^2$ which is found
to have a peak near the comoving scale $k_c$ that crosses the horizon
at the onset of transition, $k=k_{max}\sim k_c$. This in turn induces
the curvature perturbation which shows a peak near $k=k_{max}$. 
The ratio of thus induced curvature perturbation to
the conventional curvature perturbation from the inflaton field
depends on the model parameters as given in Eq.~(\ref{powers-ratio}).
For reasonable values of the parameters, we find this ratio
can become of order unity or even larger than unity.

This local feature we found may be used to explain 
the glitches found in the observed CMB angular power spectrum. 
One may also consider many waterfall fields coupled to the
inflaton to produce a series of waterfall phase transitions 
and induce multiple local features in the curvature perturbation. 
It would be interesting to study numerically the effect of single 
or multiple local features in our model and compare it with CMB data. 

Due to intrinsic non-Gaussian nature of $\delta \chi^2$ 
distribution one expects to see large spiky 
non-Gaussianities~\cite{Bond:2009xx} when 
${\cal P}_\calR^{wf}(k_{max})$ becomes comparable 
to ${\cal P}_\calR^{\phi}$.  We have shown that $f_{NL}$ has 
both intrinsic and dynamical contributions. The intrinsic
contributions originates from $\delta \chi^2$ three-point function 
while the dynamical part comes from the non-linear dynamics of
 the waterfall field. It is shown that 
$f_{NL}(k_{max}) \sim \epsilon_\chi/\epsilon C$, 
in which $\epsilon_\chi$ measures the sharpness of the waterfall 
phase transition. As a result, the sharper is the phase transition 
the larger is $f_{NL}$. As in the case of power spectrum, 
the bispectrum is highly peaked at
$k\simeq k_{max}$. It would be very interesting to investigate the 
observational consequences of these spiky non-Gaussianities.

\section*{Acknowledgement}
We would like to thank R. Allahverdi, X. Chen, E. Lim, K. Malik,  
S. Movahed, D. Mulryne and R. Tavakol  for useful discussions and comments.
 A.A.A. would like to thank ``Bonyad Nokhbegan Iran'' for partial support. 
This work was supported in part by MEXT Grant-in-Aid for 
the global COE program at Kyoto University,
"The Next Generation of Physics, Spun from Universality and Emergence,"
and by JSPS Grant-in-Aid for Scientific Research (A) No.~21244033.

\appendix

\section{Variance of waterfall field quantum fluctuations}
\label{sec:variance}

Here we calculate $\langle \delta \chi^2(n) \rangle$ in some details. 
As discussed in the text, we must consider this as part of
the classical background after it begins to evolve as a
classical field,
\begin{eqnarray}
\sqrt{\langle \delta \chi^2(n) \rangle}
\propto\exp\left[\frac{2}{3}\epsilon_\chi n^{3/2}\right]\,.
\end{eqnarray}
Then it is convenient to define $\langle \delta \chi^2(0) \rangle$
not by its actual value at $n=0$, but the value it would take
if it evolved classically from the beginning. That is,
for modes that become tachyonic by the end of the waterfall
transition $n=n_f$, we define
\begin{eqnarray}
\langle \delta \chi^2(0) \rangle\equiv
\langle \delta \chi^2(n_f) \rangle
\exp\left[-\frac{4}{3}\epsilon_\chi n_f^{3/2}\right]\,.
\end{eqnarray}
As mentioned in the text, we divide it 
into small scale and large scale parts, denoted by subscript $S$
 and $L$, respectively,
\ba
\langle \delta \chi^2(0) \rangle
= \langle \delta \chi^2(0) \rangle_S+\langle\delta\chi^2(0)\rangle_L\,.
\ea
As shown in \cite{Gong:2010zf}, for the large scale part we have
\ba
\langle \delta \chi^2(0) \rangle_L = \dfrac{H^2}{4\pi^2\epsilon_\chi}\,.
\label{dchiL2}
\ea
As for the contribution of small scale modes, as we show below, 
it was somewhat over-estimated in the previous literature.
A more accurate result is obtained as follows. 

As discussed around Eq.~(\ref{delpsi-ss-n}) in the text,
we approximately have
\ba
\delta\chi^S_k(n) = \frac{H}{\sqrt{2k}\,k_c}e^{-n_t}
\exp\left[\frac{2}{3} \epsilon_{\chi}(n^{3/2}-n_t^{3/2})\right] \,;
\quad    n> n_t(k)\,.
\label{dchiSapprox}
\ea
Literally speaking, however, the above does not reproduce the
correct behavior when $n-n_t\ll1$. At this stage, it behaves as
\ba
\delta\bar\chi^S_k(n)=\frac{H}{\sqrt{2k}\,k_c}e^{-n_t}
\exp\left[\frac{2}{3} \epsilon_{\chi}(n-n_t)^{3/2}\right]\,.
\label{dchiSapp2}
\ea
If we simply extrapolate this to $n=n_f$, we obtain
\begin{eqnarray}
\delta\bar\chi^S_k(n_f)=\frac{H}{\sqrt{2k}\,k_c}e^{-n_t}
\exp\left[\frac{2}{3} \epsilon_{\chi}(n_f-n_t)^{3/2}\right]\,,
\label{dchibarf}
\end{eqnarray}
instead of the one that follows from the approximate 
formula (\ref{dchiSapprox}),
\begin{eqnarray}
\delta\chi^S_k(n_f) = \frac{H}{\sqrt{2k}\,k_c}e^{-n_t}
\exp\left[\frac{2}{3} \epsilon_{\chi}(n_f^{3/2}-n_t^{3/2})\right] \,.
\end{eqnarray}
Then the formula (\ref{dchibarf}) gives an estimate of 
 $\delta\chi^S_k(0)$ as
\begin{eqnarray}
\delta\bar\chi^S_k(0)&=& \frac{H}{\sqrt{2k}\,k_c}e^{-n_t}
\exp\left[\frac{2}{3} \epsilon_{\chi}(n_f-n_t)^{3/2}
-\epsilon_\chi n_f^{3/2}\right]
\cr
&=&
\delta\chi^S_k(0)
\exp\left[\frac{2}{3} \epsilon_{\chi}(n_f-n_t)^{3/2}
-\epsilon_\chi(n_f^{3/2}-n_t^{3/2})\right]
\,,
\end{eqnarray}
where $\delta\chi^S_k(0)$ is estimated by using Eq.~(\ref{dchiSapprox}).
We easily see that $\delta\bar\chi^S_k(0)\leq\delta\chi^S_k(0)$.
Thus one would expect that using $\delta\bar\chi^S_k(0)$ would
give a slight underestimate, while using $\delta\chi^S_k(0)$
would give a slight overestimate. 

Let us first estimate $\langle\delta\chi^2(0)\rangle_S$
by using $\delta\chi^S_k(0)$. Using Eq.~(\ref{nt}), we find
\ba
\langle \delta \chi^2(0) \rangle_S 
= \dfrac{\epsilon_{\chi}^2 H^2}{4 \pi^2} 
\int_{n_t=0}^{n_f} dn_t \, (n_t+\dfrac{1}{2})
 e ^{-4/3 \epsilon_{\chi}n_t^{3/2}}.
\label{dchi2int}
\ea
The exponential dependence of the integrand introduces a natural 
cut-off at $n_{cut} \sim \epsilon_{\chi}^{-2/3}$. 
Hence one can neglect $n_t$ with respect to $1/2$ in the integrand,
and extend the integral range to infinity to obtain
\ba
\langle \delta \chi^2(0) \rangle_S 
&\simeq &\dfrac{\epsilon_{\chi}^2 H^2}{8 \pi^2} 
\int_{n_t=0}^{\infty} dn_t \, e ^{-4/3 \epsilon_{\chi}n_t^{3/2}} 
=\dfrac{\epsilon_{\chi}^2 H^2}{8 \pi^2}
\dfrac{\Gamma\left(\dfrac{2}{3}\right)}{6^{1/3}\epsilon_{\chi}^{2/3}}
\simeq0.75\dfrac{\epsilon_{\chi}^{4/3} H^2}{8 \pi^2}\,.
\label{delchi-var}
\ea
This gives a slightly overestimate of the true 
$\langle \delta \chi^2(0) \rangle_S$.

Now let us consider using $\delta\bar\chi^S_k(0)$.
Instead of the integral~(\ref{dchi2int}), we have
\begin{eqnarray}
\langle \delta\bar\chi^2(0) \rangle_S 
= \dfrac{\epsilon_{\chi}^2 H^2}{4 \pi^2} 
\int_{n_t=0}^{n_f} dn_t \, (n_t+\dfrac{1}{2})
\exp\left[\frac{4}{3}
\epsilon_{\chi}n_f^{3/2}\left(
\left(1-\frac{n_t}{n_f}\right)^{3/2}-1\right)\right]\,.
\end{eqnarray}
Again, similar to the previous case, the integral is
dominated by the integrand at $n_t/n_f\ll1$. Then expanding
the exponent in $n_t/n_f$, we may approximate it by
\begin{eqnarray}
\langle \delta\bar\chi^2(0) \rangle_S 
\simeq \dfrac{\epsilon_{\chi}^2 H^2}{8 \pi^2}
\int_{n_t=0}^{n_f} dn_t
\exp\left[-2\epsilon_{\chi}n_f^{1/2}n_t\right]
\simeq\dfrac{\epsilon_{\chi}^2 H^2}{8 \pi^2}
\frac{1}{2\epsilon_\chi n_f^{1/2}}
=\dfrac{\epsilon_{\chi} H^2}{16 \pi^2n_f^{1/2}}
\,.
\label{chiS2estimate}
\end{eqnarray}
This result is by a factor $\epsilon_\chi^{1/3}n_f^{-1/2}$
smaller than the estimate (\ref{delchi-var}).
For a typical value of $\epsilon_\chi$ and $n_f$, say 
$\epsilon_\chi\sim 10$ and $n_f\sim0.5$,
however, this factor is $\epsilon_\chi^{1/3}n_f^{-1/2}\sim 1.5$.
So qualitatively there is not much difference between the two
estimates.

To summarize, we conclude that the dominant contribution to
the variance of the small scale waterfall field quantum fluctuations 
comes from the modes around 
$\epsilon_\chi^{-1}n_f^{-1/2}\lesssim n \lesssim \epsilon_{\chi}^{-2/3}$,
and given by somewhere between Eqs.~(\ref{delchi-var}) and
(\ref{chiS2estimate}). Comparing these with Eq.~(\ref{dchiL2}),
we readily see that the dominant contribution to the variance
of the waterfall field quantum fluctuations comes from 
the small scale modes,
\ba
\langle \delta \chi^2(0) \rangle 
= \langle \delta \chi^2(0) \rangle_S 
+\langle \delta \chi^2(0) \rangle_L 
\simeq\langle \delta \chi^2(0) \rangle_S  \, .
\ea
Since there is not much difference between Eqs.~(\ref{delchi-var}) and
(\ref{chiS2estimate}), for definiteness we use 
Eq.~(\ref{delchi-var}) in the text.

\section{Correlation functions of square of waterfall field 
quanta ${\cal P}{\delta \chi^2}$}
\label{sec:app-2}

This appendix is devoted to find a good approximation for the 
correlation functions of the $\delta \chi^2$ appearing in the
 power spectrum and bispectrum analysis, 
Eqs.~(\ref{waterfall-conv}) and (\ref{corr-3}).

First, we work out the following convolution integral which is 
necessary is calculations of power spectrum of the waterfall field
\ba
\Big\langle \left( \delta \chi^2\right)_k 
\left( \delta \chi^2\right)_q  \Big\rangle 
&=& 2 \int \dfrac{d^3q}{(2\pi)^3} 
|\delta \chi_{|\bmk-\bmq|}|^2 |\delta \chi_q|^2 ~(2\pi)^3 
\delta^3(\bmk+\bmq).
\ea
The above integral can be divided into the radial and angular
 parts as follows
\ba
\label{int-decomp}
2 \int \dfrac{d^3q}{(2\pi)^3} |
\delta \chi_{|\bmk-\bmq|}|^2 |\delta \chi_q|^2 
= \int dn_q {\cal P}_{\delta \chi}(q) 
\int d(- \cos \theta) |\delta \chi_{|\bmk-\bmq|}|^2 \, .
\ea

The above integral can not be calculated analytically but can be 
estimated as follows. First, note that as the power spectrum of 
waterfall field ${\cal P}_{\delta \chi}$ is highly peaked around
 $k=k_{max}$ so one can expect that the peak of curvature 
perturbation ${\cal P_R} \sim {\cal P}_{\delta \chi^2}$ also take 
place near $k_{max} \gg k_c$. This assumption will be checked in 
the following. The other main point is that Eq.~(\ref{pr-delpsi}) shows 
that $|\delta \chi_{|\bmk - \bmq|} |^2$ is constant value
 $\frac{H_0^2}{2 k_c^3}$ for $|\bmk- \bmq| < k_c$ and exponentially 
decay for  $|\bmk- \bmq| > k_c$. As a result one can conclude that 
the second integral in  Eq. (\ref{int-decomp}) is negligible except for 
\ba
\label{q-cond}
|\bmk -\bmq | \lesssim \xi k_c
\ea
in which the numerical factor $\xi$ can be estimated 
from Eq. (\ref{del-psi-zero-S}). 
To find an estimate of $\xi$ it is natural to estimate the width
 of integration, $\Delta k$, the point at which $P_{\delta \chi}$ 
decreased by a factor $1/e$ from its value at $k=k_c$, 
$P_{\delta \chi}(k_c) = H_0^2/(2 k^3)$. So by using 
Eq. (\ref{del-psi-zero-S}) and noting that for 
$k \sim k_c$, $n_t(k) \ll n_{\ast} (k)$, one has
\ba
\xi \simeq e.
\ea
In order to satisfy condition (\ref{q-cond}) one can simply find
 that the amplitude of integral momentum $q$ should be near $k$ and
 at the same time the angle between $\mathbf{k}$ and $\mathbf{q}$, 
denoted by $\theta$, should be near zero. By defining
\ba
q = k+ \Delta q,\qquad \Delta q < k ,\qquad 
\mathrm{and} \qquad \theta = 0+ \Delta \theta
\ea
one can find that the condition
\ba
|\bmk -\bmq|^2 \lesssim \xi^2 k_c^2
\ea
results in
\ba
\Delta q \lesssim \xi k_c \quad 
&\longrightarrow& \quad \Delta n_q \lesssim \dfrac{\xi k_c}{k}
\\
\Delta (- \cos \theta) &=& \dfrac{\xi^2 \, k^2_c}{2k^2}
\ea
With these approximations the integral Eq. (\ref{int-decomp})
 can be estimated for $k \gg k_{c}$ as
\ba
\label{int-simp}
\nonumber
2 \int \dfrac{d^3q}{(2\pi)^3}
 |\delta \chi_{|\bmk-\bmq|}|^2 |\delta \chi_q|^2 
&=& 2 \int dn_q ~ d(-\cos \theta) 
|\delta \chi_{|\bmk-\bmq|}|^2  {\cal P}_{|\delta \chi|}(q)
\\
\nonumber
&\simeq &
\Delta n_q\,  \Delta(- \cos \theta) \dfrac{H_0^2}{2k_c^3} 
{\cal P}_{\delta \chi}(k)  
\\
&=& \dfrac{\xi^3\,H_0^2}{2k^3} {\cal P}_{\delta \chi}(k)  
\ea
Finally one can find the following good approximate for
 ${\cal P}_{\delta \chi^2}$ in terms of ${\cal P}_{\delta \chi}$ 
\ba
\label{p-chi2-app}
{\cal P}_{\delta \chi^2 }(k) 
\simeq \dfrac{\xi^3\,H_0^2}{4 \pi^2}~  {\cal P}_{\delta \chi}(k).
\ea

Similarly, we  present an approximation for the three point 
correlation function of $\delta \chi^2$ in the squeezed form
\ba
\label{conv-squeezed}
\big \langle  \left(\delta \chi^2\right)_{\bmk_1} 
\left(\delta \chi^2\right)_{\bmk_2} 
\left(\delta \chi^2\right)_{\bmk_3} 
\big \rangle_{\mathrm{sq}}
 \simeq 8 \, \delta^3(\bmk_1 + \bmk _2 + \bmk_3) \int d ^3 q 
\vert \delta \chi_q\vert^2 
\vert \delta \chi_{\vert \bmk-\bmq\vert}\vert^4 \, .
\ea
One can approximate the above integral in the same fashion as we did above
for the two point correlation function 
\ba
\big \langle  \left(\delta \chi^2\right)_{\bmk_1} 
\left(\delta \chi^2\right)_{\bmk_2} 
\left(\delta \chi^2\right)_{\bmk_3} \big \rangle_{\mathrm{sq}}
 \simeq 
8 (2 \pi)^3  \delta^3(\bmk_1 + \bmk _2 + \bmk_3) \int dn_q 
{\cal P}_{\delta \chi}(q) \,\int d(-\cos \theta) 
|\delta \chi_{|\bmk-\bmq|}|^4  
\ea
By using the above relation, one finds 
\ba
\nonumber
B^{\mathrm{sq}}_{\delta \chi^2} 
&=& 8  \int dn_q {\cal P}_{\delta \chi}(q) \,
\int d(-\cos \theta) |\delta \chi_{|\bmk-\bmq|}|^4  
\\
\nonumber
&\simeq &
8 \Delta n_q\,  \Delta(- \cos \theta) 
\dfrac{H_0^4}{4k_c^6} {\cal P}_{\delta \chi}(k)  
\\
&=& \dfrac{\xi'^3\,H_0^4}{k_c^3 k^3} {\cal P}_{\delta \chi}(k)  \, ,
\ea
in which the numerical factor $\xi'$, similarly defined as $\xi$,
 is obtained to be
 $\xi' \simeq \sqrt{e}$. Finally one has
\ba
\label{B-squeezed}
B^{\mathrm{sq}}_{\delta \chi^2}(k) 
\simeq \dfrac{\xi'^3}{k_c^3}\dfrac{H_0^4}{2 \pi^2 }\,P_{\delta \chi}(k)\,.
\ea


\section{${\delta \cal N}$  up to order $\Delta\chi^4$}
\label{sec:deltaN-fourth}

In order to find ${\delta \cal N}$  up to order $\Delta\chi^4$ 
it is enough to extend Eq.~(\ref{dpsi2}) to next order
in $\delta\chi^2$, which reads
\begin{eqnarray}
\label{dpsi-2}
\dfrac{\Delta\chi^2(n)}{\langle \delta \chi^2(n) \rangle}
-\dfrac{1}{2}
\dfrac{\Delta\chi^4(n)}{\langle \delta \chi^2(n) \rangle^2}
= \dfrac{\delta\chi_{min}^2(n_f)}{\chi^2_{min}(n_f)}
 + 2f'(n)\delta \,n - 2f'(n_f)\delta \,n_f \,.
\end{eqnarray}
This modifies $\delta {\cal N}$ in Eq.~(\ref{calN1}) to
\begin{eqnarray}
\nonumber
\delta {\cal N}
&=&\left(
\frac{\Delta\chi^2(n)}{\langle\delta \chi^2(n)\rangle}
 -\dfrac{1}{2}  \dfrac{\Delta\chi^4(n)}{\langle \chi^2(n) \rangle^2}
 \right)
\left[\dfrac{\partial n}{\partial n_f}\frac{1}{-2f'(n_f)+n_f^{-1}}\right]
\\
 &~&+\frac{\phi\,\delta\phi(n)}{2 M_P^2}
\left[1+\dfrac{C}{2}+C \epsilon n+
\frac{2f'(n)}{-2f'(n_f)+n_f^{-1}}\frac{\partial n}{\partial n_f}
\right]\,. 
\end{eqnarray}
For a sharp transition, $n_f\lesssim1$,
the above expression reduces to
\begin{eqnarray}
\label{delN-final}
\delta{\cal N}
= -\frac{C \epsilon\,n_f}{f'(n_f)}
\left(\frac{\Delta\chi^2(n)}{\langle \delta \chi^2(n) \rangle}
 -\dfrac{1}{2}\dfrac{\Delta\chi^4(n)}{\langle \chi^2(n) \rangle^2}
\right)
+\left[1+\dfrac{C}{2}+ C \epsilon(n-2n_f) \right]
\frac{\phi \delta\phi}{2 M_P^2}\,.
\end{eqnarray}
From the above, we find the relation between
$N_{,\chi^2 }$ and $N_{,\chi^2\chi^2 }$ as
\ba
\label{dN-dpsi}
N_{,\chi^2 } 
= - \langle \delta \chi^2 (0) \rangle N_{,\chi^2,\chi^2}
 = \frac{-C\, \epsilon\, n_f}{f'(n_f)} 
\frac{1}{\langle \delta \chi^2(0) \rangle}
=-\frac{C\,\epsilon}{n_f^{1/2}\epsilon_\chi}
\frac{1}{\langle \delta \chi^2(0) \rangle}\,,
\ea
where the second equality follows from the fact that
$f'(n)=\epsilon_\chi n^{1/2}$.



\end{document}